\documentclass[useAMS,usedcolumn,usenatbib,fleqn]{mn2e}

\usepackage{amsmath,amssymb}  
\usepackage{graphicx}         
\usepackage{multirow}         
\usepackage{hyperref}         
\hypersetup{colorlinks=true,linkcolor=black,citecolor=black,filecolor=black,urlcolor=black}

\newcolumntype{d}[1]{D{.}{.}{#1}}

\makeatletter
\newcommand{\extref}[1]{\textup{\tagform@{#1}}}
\makeatother

\newcommand{\unit}[1]{~\mathrm{#1}}
\newcommand{\MSun}{\mathrm{M}_\odot}
\newcommand{\RSun}{\mathrm{R}_\odot}
\newcommand{\LSun}{\mathrm{L}_\odot}
\newcommand{\LEdd}{L_\mathrm{Edd}}
\newcommand{\Teff}{T_\mathrm{eff}}

\newcommand{\Lconv}{L_\mathrm{conv}}
\newcommand{\Lrad}{L_\mathrm{rad}}
\newcommand{\Pgas}{P_\mathrm{gas}}
\newcommand{\Prad}{P_\mathrm{rad}}
\newcommand{\tKH}{t_\mathrm{KH}}
\newcommand{\Mmode}{M_\mathrm{mode}}
\newcommand{\omegar}{\omega_\mathrm{r}}
\newcommand{\omegai}{\omega_\mathrm{i}}
\newcommand{\omegaf}{\omega_\mathrm{f}}
\newcommand{\omegap}{\omega_\mathrm{p}}
\newcommand{\omegad}{\omega_\mathrm{d}}
\newcommand{\geff}{g_{\rm eff}}
\newcommand{\grad}{g_{\rm rad}}
\newcommand{\kt}{\kappa_T}
\newcommand{\kr}{\kappa_\rho}
\newcommand{\et}{\epsilon_T}
\newcommand{\er}{\epsilon_\rho}

\newcommand{\dd}{\mathrm{d}}

\newcommand{\dr}{\dd r}

\newcommand{\pd}{\partial}

\newcommand{\by}{\mathbf{y}}
\newcommand{\bA}{\mathbf{A}}
\newcommand{\bB}{\mathbf{B}}
\newcommand{\bC}{\mathbf{C}}
\newcommand{\bP}{\mathbf{P}}
\newcommand{\bxidot}{\dot\bxi}

\newcommand{\Mesa}{\textsc{Mesa}}
\newcommand{\Adipls}{\textsc{Adipls}}
\newcommand{\BV}{Brunt-V\"ais\"al\"a}

\title[Stability of metal-rich massive stars]{Stability of metal-rich very massive stars}
\author[J.\ Goodman and C.~J.\ White]{
  J. Goodman$^{1}$\thanks{E-mail: jeremy@astro.princeton.edu (JG)} and
  Christopher J.\ White$^{1}$\\
  $^{1}$Princeton University Observatory, 4 Ivy Lane, Princeton, NJ, 08544, U.S.A.}
\date{Submitted ---}

\begin{document}

\pagerange{\pageref{firstpage}--\pageref{lastpage}} \pubyear{2014}
\maketitle

\label{firstpage}

\begin{abstract}
  We revisit the stability of very massive nonrotating main-sequence stars at
  solar metallicity, with the goal of understanding whether radial pulsations
  set a physical upper limit to stellar mass. Models of up to $938$ solar masses are
  constructed with the \Mesa{} code, and their linear stability in the fundamental mode,
  assumed to be the most dangerous, is analysed with a fully nonadiabatic method. Models
  above $100\,\MSun$ have extended tenuous atmospheres (``shelves'') that affect the
  stability of the fundamental.  Even when positive, this growth rate is small, in
  agreement with previous results.  We argue that small growth rates lead to saturation at
  small amplitudes that are not dangerous to the star.  A mechanism for saturation is
  demonstrated involving nonlinear parametric coupling to short-wavelength g~modes and the
  damping of the latter by radiative diffusion.  The shelves are subject to much more
  rapidly growing strange modes.  This also agrees with previous results but is extended
  here to higher masses.  The strange modes probably saturate via shocks rather than mode
  coupling but have very small amplitudes in the core, where almost all of the stellar
  mass resides.  Although our stellar models are hydrostatic, the structure of their outer
  parts suggests that optically thick winds, driven by some combination of radiation
  pressure, transsonic convection, and strange modes, are more likely than pulsation in the
  fundamental mode to limit the main-sequence lifetime.
\end{abstract}


\section{Introduction}
\label{sec:intro}

The threshold of hydrogen burning ($\approx 0.08~\MSun$) is generally accepted as a physical lower
limit to the masses of stars, one that is independent of the environment in which stars form.
Whether there is a definite upper limit to stellar masses, and to what extent the limit may depend
on nature (stellar physics) or nurture (star-forming environment), are open questions. The highest
well-measured dynamical masses are $\sim 80~\MSun$ \citep{Schnurr2012}, most notably the
double-lined eclipsing binary WR~20a \citep{Rauw+etal2004, Bonanos+etal2004}. Statistics of stars in
galactic open clusters have been interpreted as evidence for an upper limit $\sim 150~\MSun$
\citep{Weidner+Kroupa2004,Oey+Clarke2005,Figer2005,Koen2006}, while \citet{Crowther+etal2010}
present spectroscopic arguments for larger masses among the stars in the cluster R136 of the Large
Magellanic Cloud. An empirical mass limit, if such exists, may reflect the environment in which most
stars are observed to form: that is to say, molecular clouds, where the density of hydrogen
nuclei is typically $n \lesssim 10^{3}\unit{cm^{-3}}$, the temperature $\lesssim 100\unit{K}$, and
dust is abundant.

One of us has previously argued that the broad-line regions of bright QSO accretion disks
are likely self-gravitating and prone to form very massive stars -- at least several
hundred solar masses at the onset of gravitational instability, and perhaps
$\ga 10^5~\MSun$ after accretion up to the isolation mass
\citep{Goodman+Tan2004,Jiang+Goodman2011}.  A QSO disk at $\sim 10^3$ gravitational radii
from the black hole is a very different environment from a molecular cloud: denser by many
orders of magnitude, hotter than the sublimation temperature of dust, rapidly rotating and
shearing, and dominated by radiation pressure rather than gas pressure. Hence a different
initial-mass function and maximum stellar mass might result in such disks than in giant
molecular clouds. On the other hand, it is well known that very massive stars are fragile
due to the predominance of radiation over gas pressure, radiatively driven winds, and
pulsational instabilities. Thus it is possible that internal physics establishes an upper
limit $\lesssim 10^2\text{--}10^3~\MSun$. A presumably fatal relativistic instability sets
in above $10^5\text{--}10^6~\MSun$, depending upon internal rotation
(\citealt{Chandrasekhar1964,Baumgarte+Shapiro1999,Montero+Janka+Mueller2012}, and
references therein). This leaves a gap of several orders of magnitude above the largest
observed masses, however.

In the present paper, we return to the question of pulsational instabilities driven by the $\kappa$-
and $\epsilon$-mechanisms, which are sensitive to composition via opacities and to nuclear reaction
rates. This is a problem that has been considered by many authors since the original work by
\citet{Schwarzschild+Haerm1959}, and one might have thought it a closed subject. However, the
understanding of the opacities and other microphysical inputs has evolved, while the effects of
convection on the linear growth rates remain uncertain, as do the mechanisms responsible for
nonlinear saturation of the pulsations if they grow at all.

We focus on the fundamental radial mode, on the assumption that it is most dangerous.
\cite[hereafter GK93]{Glatzel+Kiriakidis1993} have analyzed the stability of solar-metallicity stars up to
$120\,\MSun$ and found a host of higher-order modes that grow more quickly than the fundamental.
However, the energies of these modes are concentrated in the outer parts of the star, so that they
can be expected to reach nonlinear amplitudes before the bulk of the mass is much affected.  The
fundamental involves the entire star.  Its character and scaling with mass are unlike those of other
modes.  The pulsation period increases $\propto M^{1/2}$, whereas those of higher-order radial modes
scale $\propto M^{1/4}$.
This is due to the predominance of radiation pressure, $\Prad/\Pgas\propto M^{1/2}$ for
$M\gtrsim 100\,\MSun$, which makes very massive nonrotating stars almost neutrally stable
against changes in radius even in the adiabatic approximation.
Thus at large amplitudes ($\delta r/r\gtrsim 1$), the fundamental mode might
disrupt or collapse the entire star.  The former requires unbinding the star,
which becomes less difficult with increasing mass because of the predominance of
radiation pressure and the attendant near-cancellation of gravitational and potential
energies in hydrostatic equilibrium.  (Indeed, spontaneous disruptions occurred in the
simulations of AGN disk fragmentation by \cite{Jiang+Goodman2011}, though they were
caused by numerical energy errors rather than pulsations.)  Collapse might occur under
extreme compression due to electron-positron pairs at central temperatures approaching
$10^9\unit{K}$, or due to relativistic corrections to gravity at masses $>10^4\MSun$
\citep{Zeldovich+Novikov1971}.

Recently, \citet[hereafter SQA]{Shiode+Quataert+Arras2012} have revisited the $\epsilon$-mechanism.
Applying a quasi-adiabatic analysis to equilibrium models constructed with the \Mesa{} code
\citep{Paxton_etal2011,Paxton_etal2013}, they concluded that the instability is suppressed by the
effective viscosity due to turbulent convection, at least for stars of masses $\la 1000~\MSun$.
However, they did not consider any models above $100~\MSun$ with solar or higher metallicity. Since
QSO disks appear to be metal rich, with metallicities perhaps up to ten times solar
\citep{Hamann+Ferland1999,Dietrich+etal2003,Matsuoka+etal2011,Batra2014}, one motivation for the
present work was to repeat SQA's analysis at higher metallicities \emph{and} masses $> 10^2~\MSun$.
We also wanted to perform a fully non-adiabatic rather than quasi-adiabatic analysis. This is
arguably less important for the $\epsilon$-mechanism because it is driven deep within the star where
the thermal time is very long. However, SQA also found evidence for instabilities driven by opacity
variations in the envelope which they did not fully explore, perhaps because they had less
confidence in the quasi-adiabatic approximation for those modes. Also, at least with modern
opacities, the envelopes of high-mass stellar models at solar metallicity differ strikingly from
those of corresponding Population~III models, and this has interesting consequences for the mode
structures.

Linear stability analysis is only a first step toward answering the question posed above. If
instabilities are found, one must consider how they may saturate in order to decide whether they are
likely to shorten the main-sequence lifetime. Early attempts to address the saturation of
instabilities driven by the $\epsilon$-mechanism gave conflicting results
\citep{Appenzeller1970,Papaloizou1973b}, but little work has been done along these lines in recent
decades. We will argue that even if the uncertain damping effects of convection are neglected, the
linear growth rates are so small compared to the real part of the pulsation frequency that the
pulsations will saturate by one or another weakly nonlinear mechanism at small amplitudes that do
not threaten the survival of these stars, at least not before they have lived out most of the
nominal minimum main-sequence lifetime ($\sim 3\times 10^6\unit{yr}$). Instead, in view of the
structure of our hydrostatic models, as well as a recent body of work
on Wolf-Rayet and O-star winds, radiatively driven mass loss seems more likely
likely than pulsational instabilities to limit the lifetimes of the most massive, metal-rich stars.
We hope to explore the scaling of the mass-loss timescale (i.e., $\lvert M/\dot{M}\rvert$) with
stellar mass in a future paper.

The outline of this paper is as follows.  Section~2, supplemented by an Appendix, presents
the equilibrium \Mesa{} models and our methods for the linear stability analysis.
Section~3 highlights the extended atmosphere or `shelf' seen in the higher-mass models.
Results for the growth rates and eigenfunction of the fundamental are given in \S4, with
particular emphasis on nonadiabatic effects in the shelf.
\cite{Glatzel+Kiriakidis1993}'s intrinsically nonadiabatic `strange modes' are
shown to extend to higher masses, where they have longer periods than the fundamental.
\S5 examines nonlinear saturation of the fundamental through 3-mode or parametric coupling
to high-order nonradial g-modes, and (more briefly) saturation of strange modes
in shocks.  Since the stably-stratified zones of our most massive models are relatively
small, and would perhaps disappear entirely at some higher mass, the explicit estimates in
\S5 are intended to be illustrative of a larger class of weak nonlinearities that will
limit the amplitude of the fundamental when its growth rate is small.  A summary of our
conclusions and a discussion of future steps follows in \S6.

\section{Method}
\label{sec:method}

\subsection{Equilibrium Models and Initial Estimates}
\label{sec:method:setup}

Like SQA, we generated ZAMS stellar models using the stellar evolution code \Mesa{}. Here we
highlight the configuration settings used when they deviate from the defaults.

\begin{table}
  \centering
  \caption{Basic properties of our ZAMS models}
  \label{tab:models}
  \begin{tabular}{rrrrr} 
    \hline
    \multicolumn{1}{c}{$M/\MSun$} &\multicolumn{1}{c}{$L/\LSun$} &
    \multicolumn{1}{c}{$R/\RSun$} & \multicolumn{1}{c}{$T_{\rm eff}$ [K]} &
    \multicolumn{1}{c}{$X_{\rm c}$}\\ \hline
                           10.0 &$5.118\times10^3$ & 3.922& 24688 &  0.727 \\
                           21.5 &$4.841\times10^4$ & 5.998& 34983 &  0.727 \\
                           46.4 &$2.962\times10^5$ & 9.279& 44236 &  0.727 \\
                           100. &$1.244\times10^6$ & 15.27& 49360 &  0.729 \\
                           215. &$4.054\times10^6$ & 30.88& 46638 &  0.726 \\
                           464. &$1.135\times10^7$ & 75.29& 38635 &  0.728 \\
                           938. &$2.686\times10^7$ & 274.3& 25106 &  0.727 \\ \hline
  \end{tabular}
\end{table}

\begin{figure}
  \centering
  \includegraphics[width=\columnwidth]{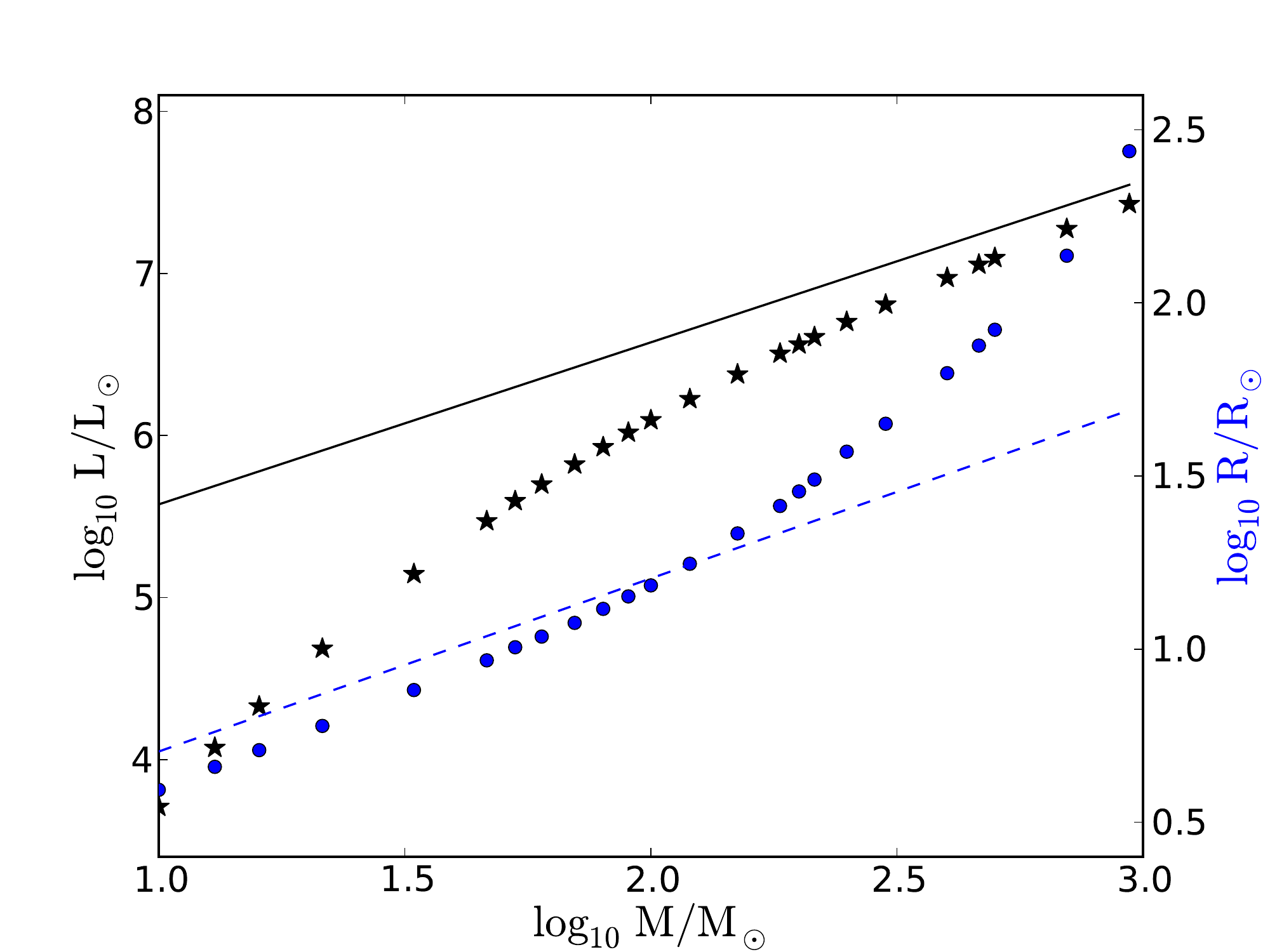}
  \caption{Luminosity (star symbols, left scale) and photospheric radius (circles, right
    scale) versus stellar mass for our \Mesa{} models. Thin solid line is the Eddington
    limit for this composition ($Y=0.25$, $Z=0.02$).  Thin dashed line shows
    $R\propto M^{1/2}$, as would be expected for homologous radiation-pressure-dominated
    models.}
  \label{fig:LandR}
\end{figure}

The initial mass is specified, with initial abundances
$(X,Y,Z) = (0.73,0.25,0.02)$.  The simulation begins in the pre-main sequence phase
at a large radius and low central temperature.  The atmosphere is modelled as a `simple
photosphere.' Convective mixing is implemented following
\citet{Henyey+Vardya+Bodenheimer1965} in regions determined to be convectively unstable by
the Ledoux criterion.

The star is evolved until it is determined to lie on the main sequence, defined by the
minimum photospheric radius.  (\Mesa{} has its own way of deciding when the model has
reached the main sequence, but we found its decisions unreliable for our higher-mass
models.)  The nuclear and photospheric luminosities are then in equilibrium, and the
central hydrogen abundance is only very slightly depleted.  Table~\ref{tab:models} lists
some properties of the models, which are similar to those of GK93 (within $5\%$ in $L$ and
$1\%$ in $T_{\rm eff}$), except that theirs were limited to 40-120~$\MSun$.  The effective
temperature peaks at $4.95\times10^4\unit{K}$ near $120\MSun$: more massive are cooler
because of their distended atmospheric ``shelves'' driven by iron opacities (\S3).

Once the star has reached its ZAMS phase, the model is saved and the data is analysed with the
\Adipls{} package \citep{CDalsgaard2008}, which computes adiabatic pulsational modes given
one-dimensional stellar models.
The output from \Adipls{} is a set of eigenfrequencies and corresponding mode shapes in
the form of radial displacements from equilibrium. We use these outputs in our own
routine, which finds mode frequencies and shapes without the adiabatic assumption. We turn
to this method now.

\subsection{Nonadiabatic Analysis}
\label{sec:method:nonadiabatic}

For the purpose of analysing the pulsational modes in a nonadiabatic framework, we began
by adopting the method outlined in \citet{Castor1971}, with mass fraction as the
independent variable, and a Henyey-type relaxation scheme for finding the
eigenfrequencies.  Despite extensive efforts and algorithmic variations, this did not give
numerically stable results for the growth rate of the fundamental mode, possibly because
of the many orders of magnitude separating the dynamical and thermal timescales in the
core, and the enormous radial variation in the ratio of these timescales through the star.

We therefore adopted a sort of shooting method designed for very stiff equations (Appendix
\ref{sec:stiff}).  The basic linearized equations are
\begin{subequations}\label{eq:laglin}
  \begin{align}
    \label{eq:d1}
    \frac{\pd}{\pd r} \delta\Pgas &= \rho\left[\omega^2\delta r + \grad
      \left(\frac{\delta\kappa}{\kappa}+\frac{\delta \Lrad}{\Lrad}\right) +
          4\geff\frac{\delta r}{r}\right],\\
     \label{eq:d2}
     r^{-2}\frac{\pd}{\pd r}(r^2\delta r) &= -\frac{\delta\rho}{\rho},\\
     \label{eq:d3}
     \frac{\pd}{\pd r}\delta\Prad &=
     -\rho\grad\left(\frac{\delta\kappa}{\kappa}+\frac{\delta\Lrad}{\Lrad}-4\frac{\delta
         r}{r}\right),\\
     \label{eq:d4}
     \frac{\pd}{\pd r}\delta L &= i\omega 4\pi r^2\rho T\delta S + 4\pi r^2\rho\frac{\delta\epsilon}{\epsilon}\,.
  \end{align}
\end{subequations}
Here $\grad =\kappa\Lrad/4\pi r^2 c$ is the radiative force per unit mass, $\geff =
GM_r/r^2\,-\grad$ is the residual between the gravitational and radiative accelerations, $\delta$
represents lagrangian perturbation (first-order variation at fixed interior mass), and all other
symbols have their usual meaning.
We define dimensionless linearized variables
\begin{equation}
  \label{eq:ydefs}
  y_0\equiv\frac{\delta r}{r}\,,\quad y_1\equiv\frac{\delta\rho}{\rho}\,, \quad y_2\equiv\frac{\delta
    T}{T}\,,\quad y_3\equiv\frac{\delta\Lrad}{L}\,.
\end{equation}
Notice that $L$ not $\Lrad$ appears in the denominator of $y_3$.  

In principle $\delta L = \delta\Lrad + \delta\Lconv$.  However, since there is no
generally accepted prescription for time-dependent convective
luminosity---especially in the radiation-pressure-dominated regime---we adopt
$\delta\Lconv=0$ (`frozen convection').  Nor have we allowed for a turbulent convective
viscosity in the linearized momentum equation \eqref{eq:d1}.

Then in terms of our dimensionless variables, with primes for $d/dr$ and writing
$f\equiv\Lrad/L$, $\kt=(\partial\ln\kappa/\partial\ln T)_\rho$, $\kr=(\partial\ln\kappa/\partial\ln
\rho)_T$,
and similarly for $\et$ and $\er$, eqs.~\eqref{eq:laglin} become
\begin{subequations}\label{eq:dydr}
\begin{align}
ry_0' &= -3y_0-y_1\\
y_1'+y_2' &= \frac{\rho}{\Pgas}[\omega^2r y_0+\grad(\kr y_1 +\kt
  y_2+f^{-1} y_3)\nonumber\\
        &\quad +\geff (4y_0+y_1+y_2)]\\
y_2' &= -\frac{\rho\grad}{4\Prad}[\kr y_1+\kt y_2 +f^{-1}y_3-4y_0-4y_2]\\
ry_3' &= \frac{4\pi r^3\rho}{L}\{i\omega C_VT[y_2-(\Gamma_3-1)y_1]\nonumber\\
       &\quad +\epsilon(\er y_1 +\et y_2 - y_3)\}\,.
\end{align}
\end{subequations}

The system of equations is closed by choosing four boundary conditions.  Physically, one
expects $\delta r = \delta L = \delta\Lrad =0$ at $r=0$.  This does not require $y_0$ or
$y_3$ to vanish at the centre, but from the first of eqs.~\eqref{eq:dydr}, one sees that
nonsingular behaviour requires
\begin{equation}
  \label{eq:bcin1}
  3y_0+y_1\to0\mbox{ as }r\to0.
\end{equation}
Similarly, regularity of the last of eqs.~\eqref{eq:dydr} implies
\begin{equation}
  \label{eq:bcin2}
  \frac{i\omega C_VT}{\epsilon}[y_2-(\Gamma_3-1)y_1]+ \er y_1 +\et y_2 - y_3\to 0\mbox{ as }r\to0.
\end{equation}
The factor in front of the square brackets is $\sim\tKH/t_{\rm dyn}\gg 1$ if $\omega\sim t_{\rm
  dyn}^{-1}$, so to a first approximation the behaviour near the origin is adiabatic, $\delta\ln
T\approx (\Gamma_3-1)\delta\ln\rho$.  But since we are interested in growth or decay rates
$\omega_I\sim\tKH^{-1}$, we use eq.~\eqref{eq:bcin2} as written.

The outer heat equation requires a more general analysis than is given by \citeauthor{Castor1971},
as the equations in that work only hold under the assumption that radiation pressure is negligible
compared to gas pressure at the outer boundary. This condition clearly does not hold for very
massive stars. We therefore turn to the equation for radiation pressure in the Eddington
approximation,
\begin{equation*} 
  \Prad = \frac{F}{c} \left(\tau + \frac{2}{3}\right)\text{,}
\end{equation*}
where $F$ is the radiative flux and $\tau$ is the optical depth at the location being considered. In
more familiar variables, 
\begin{equation}\label{eq:prad}
  \Prad = \frac{\Lrad}{4\pi r^2c} \left(\frac{\kappa\Delta m}{4\pi r^2} + \frac{2}{3}\right),
\end{equation}
where $\Delta m$ is the mass exterior to the point being considered, if
the density scale height is $\ll r$.
Linearizing yields
\begin{equation*}
  \frac{\delta\Prad}{\Prad} = \frac{\delta\Lrad}{L} - 4 \left(\frac{\tau+1/3}{\tau+2/3}\right)
  \frac{\delta r}{r}
 + \left(\frac{\tau}{\tau+2/3}\right) \frac{\delta\kappa}{\kappa},
\end{equation*}
or in dimensionless variables,
\begin{equation}
  \label{eq:bcout1}
  -4(\tau+\tfrac{1}{3})y_0 +\tau\kr y_1 + \tau\kt y_2 +(\tau+\tfrac{2}{3})(y_3-4y_2)=0.
\end{equation}
We replace $\Lrad$ with $L$ here because in practice the model extends far enough into the tenuous
atmosphere as to make the contribution of $\Lconv$ negligible.  

The outer momentum equation is closed as follows. The total pressure at a point near the surface,
under a mass $\Delta m$ and at a radius $r$, is
\begin{equation} \label{eq:ptot}
  P = \frac{\Delta m}{4\pi} \left(\frac{\ddot{r}}{r^2} + \frac{GM}{r^4}\right) + \frac{L}{6\pi cr^2},
\end{equation}
if the mass shell $\Delta m$ is thin and effectively hydrostatic in its accelerated frame.  The last
term in \eqref{eq:ptot} is the radiation pressure extrapolated to the outer surface of the shell,
where $\tau=0$.
Subtracting \eqref{eq:prad} from \eqref{eq:ptot} yields
\begin{equation}
  \Pgas = \frac{\Delta m}{4\pi} \left(\frac{\ddot{r}}{r^2} + \frac{GM}{r^4} - \frac{\kappa\Lrad}{4\pi r^4c}\right)\text{.}
\end{equation}
Perturbing this yields
\begin{equation*}
  \frac{\delta\Pgas}{\Pgas} = -\frac{\delta r}{r} \left(4 + \frac{\omega^2r^3}{\beta GM}\right) - \frac{1-\beta}{\beta} \left(\frac{\delta L}{L} + \frac{\delta\kappa}{\kappa}\right),
\end{equation*}
or equivalently,
\begin{equation}
  \label{eq:bcout2}
   \left(4 + \frac{\omega^2 R^3}{\beta GM}\right)y_0 + \frac{1-\beta}{\beta}(y_3 + \kr y_1 + \kt y_2)
 -y_1-y_2 = 0.
\end{equation}
where $\beta$ is defined in terms of the Eddington luminosity $\LEdd = 4\pi GMc/\kappa$ by $L =
(1-\beta) \LEdd$.

\section{The shelf}
\label{sec:atmosphere}

\begin{figure}
  \centering
  \includegraphics[width=\columnwidth]{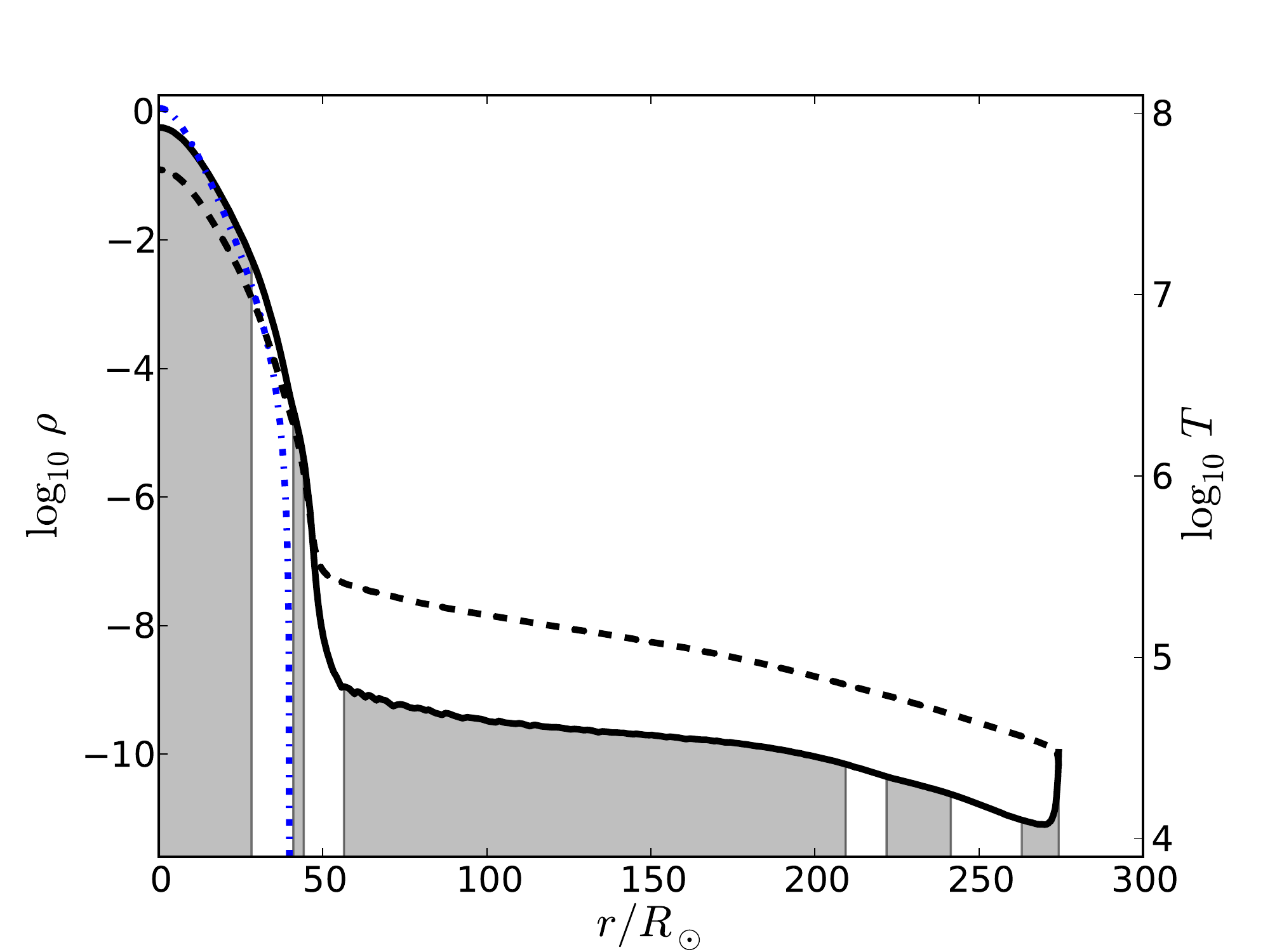}
  \caption{Density (solid line, left ordinate) and temperature (dashed line, right
    ordinate) in our $938~\MSun$ model.  Grey shading indicates convective regions.
    Dot-dashed line is the density profile of an $n = 3$ polytrope scaled according to
    eq.~\extref{9} of \citet{Goodman+Tan2004}.\label{fig:shelf} }
\end{figure}

When \Mesa{} evolves massive stars with nonnegligible metallicity, it generically produces an
extended envelope outside the polytropic core of the star, as shown in
Figure~\ref{fig:shelf}. This extremely diffuse region, which
is incipient in the $46~\MSun$ model but prominent in those above $100~\MSun$, occupies a
progressively larger fraction of the star's radius but a minute fraction of its mass
($\Delta M_{\rm shelf}\approx 3\times10^{-6}M$ for the $938\MSun$ model). It is
overwhelmingly dominated by radiation pressure, much more so than the stellar core.  Because of its
slowly radially varying temperature and density, we call this region the `shelf,' although the
density profile is actually inverted in its outermost part (Fig.~\ref{fig:shelf}). The
opacity in the shelf rises above the electron-scattering value. Since hydrostatic equilibrium limits
the radiative part of the luminosity to the value that just balances gravity, $\Lrad = 4\pi
GM_*c/\kappa(\rho, T)$, and since the luminosity is near-Eddington, the balance of the luminosity
(up to half, in our most massive model) is carried by inefficient convection that approaches the
adiabatic sound speed, i.e. the sound speed based on total rather than gas pressure.  A similar
shelf has been observed in models of Wolf-Rayet stars and attributed to a bump in the iron opacity
at temperatures $\sim\mbox{1-2}\times10^5\mathrm{K}$ \citep[and references
therein]{Grafener+Owocki+Vink2012}.  

\Mesa{} uses OPAL radiative opacities \citep{Iglesias+Rogers1996} in the density and
temperature regime relevant to the shelf.  We employ the so-called Type~1 opacities
(carbon and oxygen abundances not determined independent of metallicity) with ``solar''
relative abundances as defined by \citet{Grevesse+Noel1993a}.  See
\citet{Paxton_etal2011}, \S4.3, for more details.

\begin{figure}
  \centering
  \includegraphics[width=\columnwidth]{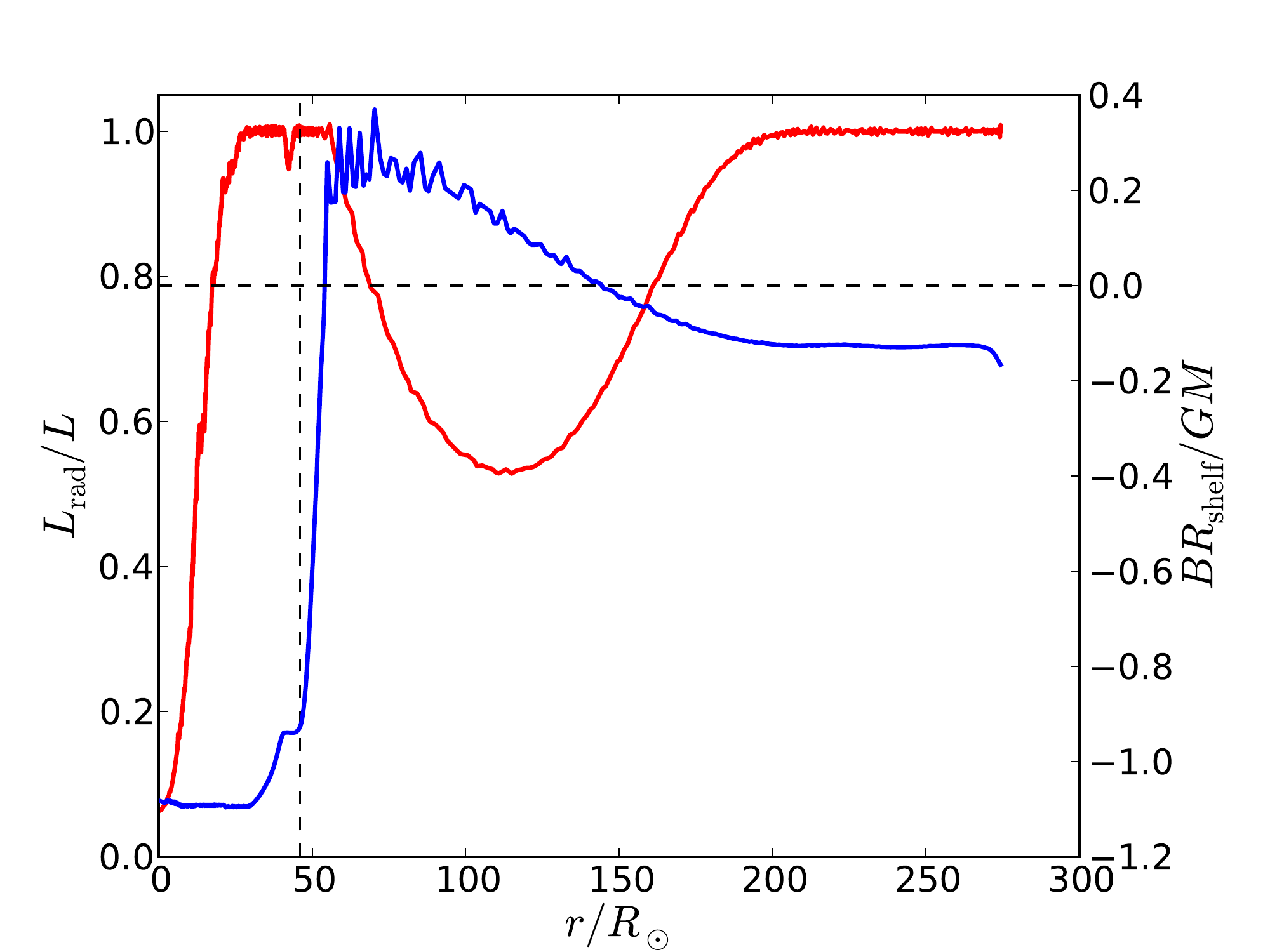}
  \caption{{\it Left axis, red curve:} Fraction of luminosity carried by radiative diffusion in the
    $938\MSun$ model. {\it Right axis, blue curve:} Bernoulli ``constant'' normalized by mass and
    radius at the base of the shelf.}
  \label{fig:Lfrac}
\end{figure}

In fact, the shelf might be
replaced by a radiatively driven wind if the constraint of hydrostatic equilibrium were
relaxed. This is suggested by the fact that the Bernoulli `constant'
\begin{equation}
  B = u + \frac{1}{2} v^2 + \frac{P}{\rho} + \Phi
\end{equation}
becomes positive in the lower part of the shelf, though it changes sign once more in the outer
convective regions. Here $u$ is the internal energy per unit mass, and
$v$ is the \emph{mean} radial velocity, which of course vanishes in these hydrostatic
models, and $\Phi(r)$ is the gravitational potential, defined to vanish as $r\to\infty$. The
kinetic energy of the convection would further increase $B$.  These regions being strongly
non-adiabatic, however, $B>0$ does not guarantee a successful wind.

For a precise definition of the base of the shelf, we use the radius or
mass fraction corresponding to the local minimum in the pressure scale height, $H_P$.  For the model
shown in Fig.~\ref{fig:shelf}, this is $R_{\rm shelf}=46.3\,\RSun$.  Within our suite of models,
such a minimum occurs only for $M\gtrsim 50\,\MSun$.

\section{Results}
\label{sec:results}

\subsection{Adiabatic Calculations}
\label{sec:results:adiabatic}

\begin{figure}
  \centering
  \includegraphics[width=\columnwidth]{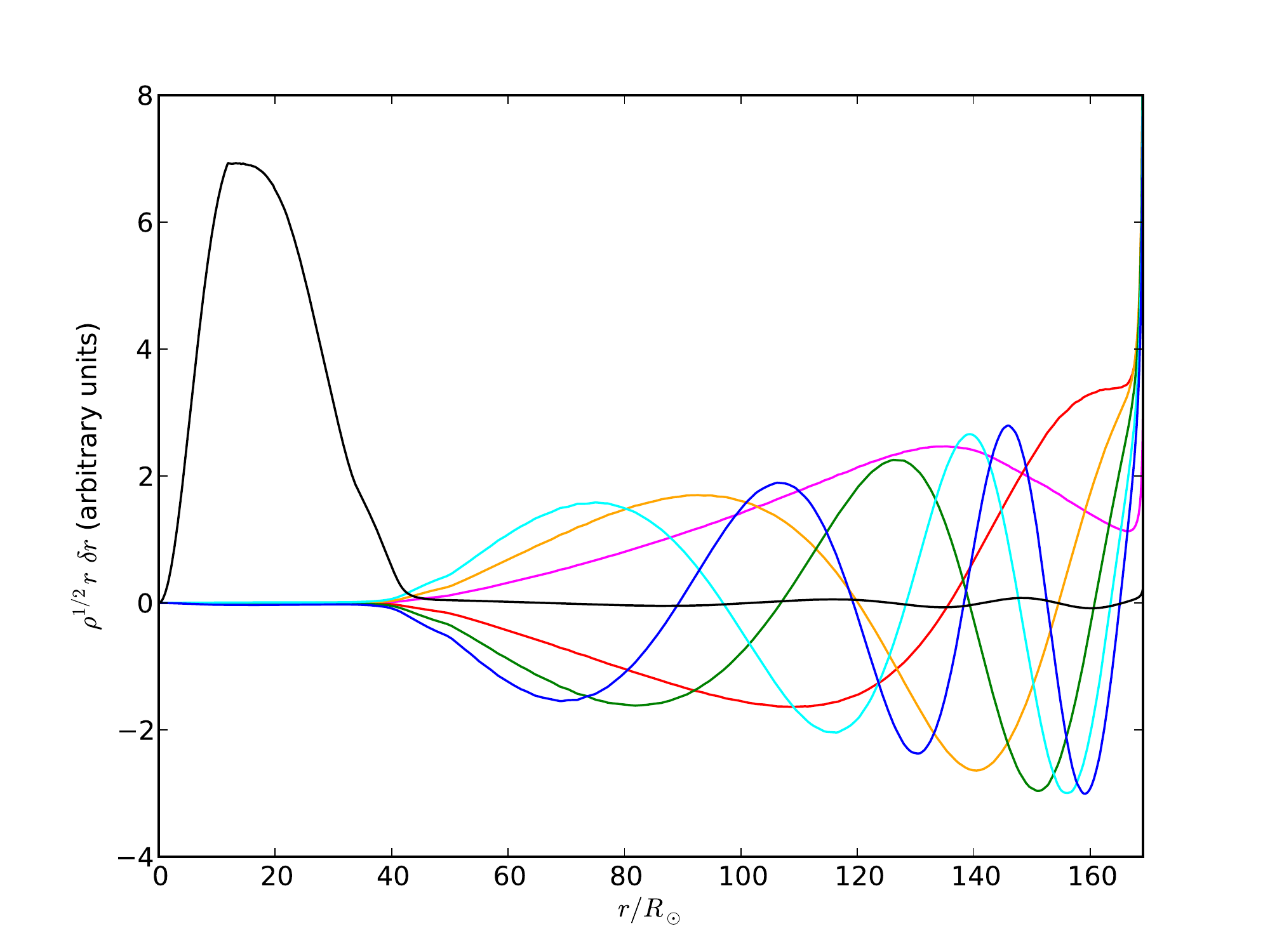}
  \caption{Plot of the lowest frequency adiabatic modes for the $464~\MSun$ model. The curves show
    the square root of the mode kinetic energy per unit length, normalized independently. The modes
    have radial mode number $n$ equal to $1$ (red), $2$ (orange), $3$ (green), $4$ (cyan), $5$
    (blue), $6$ (magenta), or $7$ (black), only the latter of which is not evanescent in the core.
    \label{fig:adiabatic_modes}}
\end{figure}

First we describe the results obtained using \Adipls. The seven lowest-frequency modes for our
$464~\MSun$ model are shown in Figure~\ref{fig:adiabatic_modes}. Weighting the displacement $\delta
r$ by $\rho^{1/2} r$ shows where the energy of the mode is concentrated, in that $\omega^2$ times
the integral of the square of the plotted quantity gives the total energy.

Of particular note is that the first six modes discovered by \Adipls{} are trapped in the shelf.
The seventh mode is the true fundamental: its energy is concentrated in the core, where it has no
nodes other than the centre, and it has an antinode at the base of the shelf (see
\S\ref{sec:atmosphere}).  Some readers may object to our use of the term ``fundamental'' for a mode
that has multiple radial nodes.  The adiabatic linearized problem is of Sturm-Liouville type, with
orthogonal eigenfunctions having interleaved nodes.  But the nonadiabatic problem is quite
different.  As Figure~\ref{fig:eig938} and Table~\ref{tab:growth_rates} illustrate, the adiabatic
and non-adiabatic versions of the fundamental (as we define it) are very similar---and both
nodeless---in the core, and have nearly equal real parts of their eigenfrequencies, but can have
different numbers of nodes in the shelf (11 and 8, respectively, for the real parts of the
eigenfunctions shown in Fig.~\ref{fig:eig938}).  Thus classification on the basis of numbers of
nodes is not helpful in establishing a correspondence between the adiabatic and non-adiabatic
eigenfunctions.  Perhaps some other term such as ``basic'' could be substituted for ``fundamental,''
but we feel that the latter is physically justified in this application.

\Adipls{} also reports the frequencies of these modes, which are of course real in the adiabatic
approximation.  The corresponding periods, $2\pi/\omega$, of the fundamental modes differ only in
the third or fourth significant digit from the values shown in the second column of
Table~\ref{tab:growth_rates} for the fully nonadiabatic fundamental modes.

\subsection{Nonadiabatic Calculations}
\label{sec:results:nonadiabatic}

In the linear equations, nonadiabaticity arises from two primary mechanisms. The nuclear heating
rate per unit mass, $\epsilon$, is sensitive to density and even more so to temperature.  The
strongly positive value of the logarithmic temperature derivative $\epsilon_T$ ($\approx 12$ near the
centre of the $938\MSun$ model) tends to add entropy during the compressive phase of the pulsation
cycle when $\delta T>0$, thus producing mechanical work.  This is the classic epsilon mechanism.
The entropy of mass elements varies also by radiative diffusion.  This occurs in the linear analysis
even if the opacity is constant, due to perturbations in the temperature gradient, but instability
by the kappa mechanism generally requires that $\kappa_T>0$ in regions of the star where the local
thermal time $t_{\rm th}\equiv L^{-1}C_V H_P dM_r/dr$ is comparable to the pulsation period (e.g.,
\citealt{Cox1980}).  Convection may tend to stabilize pulsations by providing an effective
viscosity, but as discussed by SQA and references therein, the viscous effect is thought to
be suppressed when the convective turnover time is long compared to the pulsation period.
It is also possible for convection to drive instability when it adjusts rapidly to the changing
superadiabatic gradient \citep{Brickhill1991}.  Except for a quasi-adiabatic estimate along
the lines of SQA, we have generally neglected these convective effects, even though these stars
are in fact largely convective.

When applying the numerics described in \S\ref{sec:method:nonadiabatic}, we are free to set any of
$\epsilon_T$, $\epsilon_\rho$, $\kappa_T$, and $\kappa_\rho$ to zero throughout the model. In this
way we can separate the two effects.  Thus the third column of Table~\ref{tab:growth_rates} lists
the growth rates obtained when the derivatives of $\kappa$ are neglected, and similarly the fourth
column gives the rates when the derivatives of $\epsilon$ are set to zero, while the fifth column
retains all derivatives. 

The effects of the nonadiabatic mechanisms on the growth rate are not entirely additive,
as they would be in the quasi-adiabatic approximation.  This is because the opacity
derivatives have a substantial effect on the shape of the eigenfunction near the
photosphere (or in the shelf), which is also the region mainly responsible for driving or
damping.  However, the epsilon mechanism does appear to be additive, as might be expected
since it acts only in the core where the quasiadiabatic approximation is excellent.  That
is to say, if $\omega_I^{(\epsilon)}$, $\omega_I^{(\kappa)}$, and $\omega_I$ represent the
growth rates in the third through fifth columns of Table~\ref{tab:growth_rates}, while
$\omega_I^{(0)}$ is the growth rate obtained when all of $\epsilon_T$, $\epsilon_\rho$,
$\kappa_T$, $\kappa_\rho$ are neglected (this is not shown in the Table), then we do find
that $\omega_I^{(\epsilon)}-\omega^{(0)}\approx\omega_I-\omega_I^{(\kappa)}$ for
all of the models except perhaps the first ($10\MSun$), in which the
$\epsilon$ mechanism is very weak.

\begin{table}
  \centering
  \caption{Periods and growth rates of fundamental radial mode. Negative growth rates indicate
stability.  Note $1\,\mathrm{Md}\equiv 10^6\,\mathrm{day}$.
\label{tab:growth_rates}
}
  \begin{tabular}{ccccccc}
    \hline
    \multicolumn{1}{c}{$M$} &\multicolumn{1}{c}{Period} &
    \multicolumn{5}{c}{Growth Rate ($\mathrm{Md}^{-1}$)} \\ \cline{3-7}
    \multicolumn{1}{c}{($\MSun$)}&\multicolumn{1}{c}{(d)}& \multicolumn{1}{c}{$\epsilon$} &
                       \multicolumn{1}{c}{$\kappa$} & \multicolumn{1}{c}{total} &
                       \multicolumn{1}{c}{convec.} &\multicolumn{1}{c}{no shelf}\\ \hline
                           10    &0.0905& -22.4  &   41.0 &   41.1 &   41.1  & ---    \\
                           21.5 &0.1391& -6.94  & -10.3 & -9.64 & -9.68  & ---    \\
                           46.4 &0.2172& -3.01  & -7.93 & -6.33 & -6.50  & ---    \\
                           100  &0.3279& -3.12  & -2.37 & +0.44 & 0.015  & -1.73 \\
                           215  &0.4947& -19.5  & -5.15 & +9.28 & 8.45    & +1.22 \\
                           464  &0.7253& -5.49  & -11.1 & -5.92 & -7.19  & +3.26 \\
                           938  &1.0435& -9.87  & -15.5 & -10.6 & -11.9  & +3.69 \\ \hline
  \end{tabular}
\end{table}

The sixth column in Table~\ref{tab:growth_rates} differs from the fifth by including a
quasi-adiabatic work-integral estimate of convective viscous damping, following
equations~\extref{11} and~\extref{12} of SQA.  Since the quasi-adiabatic method depends
upon the adiabatic eigenfunction, and since the adiabatic and nonadiabatic eigenfunctions
differ strongly in the shelf region, we truncate the work integrals at the local minimum
in the pressure scale height.

\begin{figure}
  \centering
  \includegraphics[width=\columnwidth]{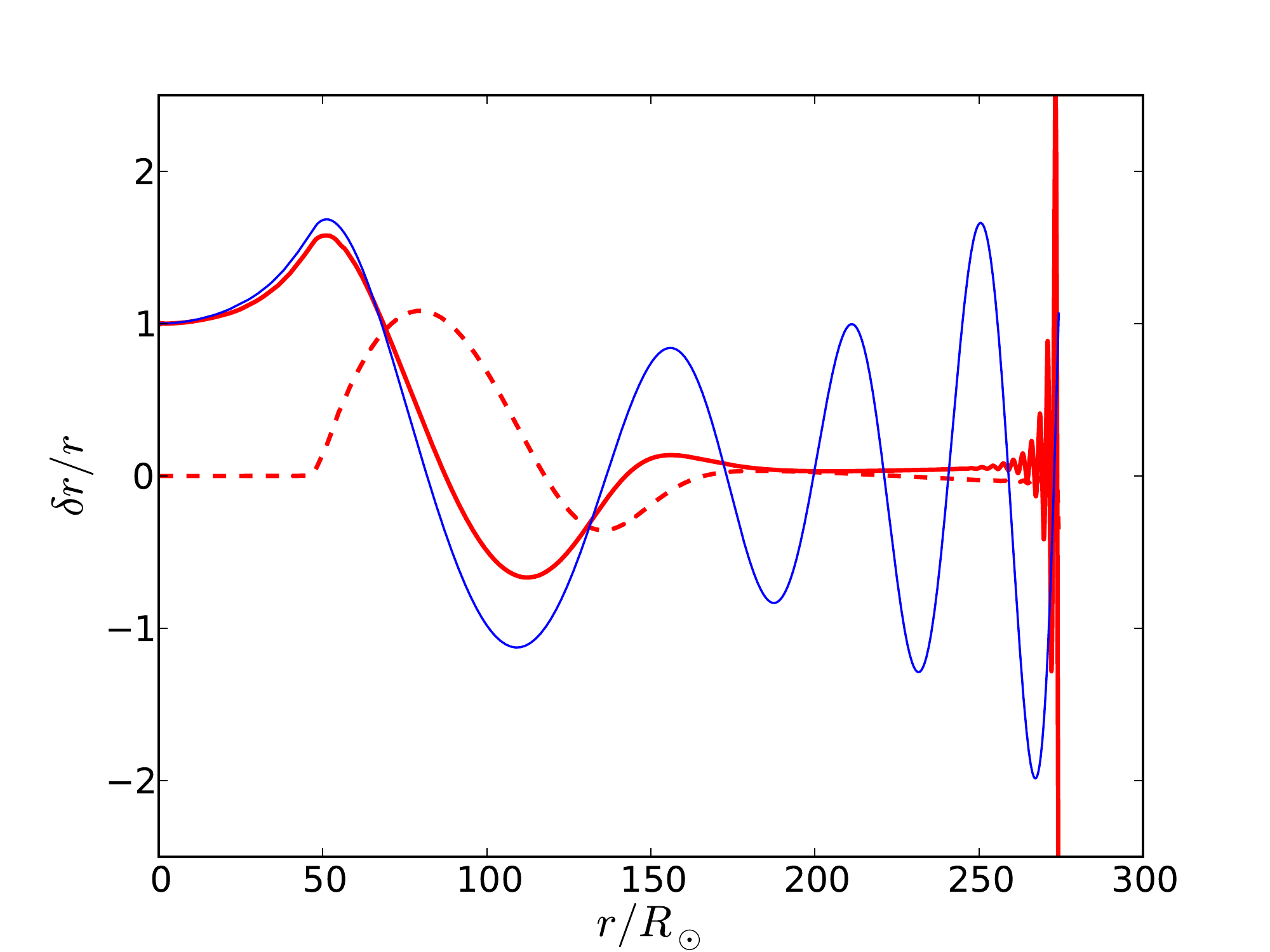}
  \caption{Fundamental radial mode at $938\MSun$.  {\it Thin blue line:} Adiabatic
    displacement eigenfunction. {\it Red lines:} Real (solid) and imaginary (dashed) parts
    of the fully nonadiabatic displacement eigenfunction.  Eigenfunctions are not
    multiplied by $r^2\rho^{1/2}$, in order to emphasize behaviour in the shelf
    ($r>46\RSun$).}
  \label{fig:eig938}
\end{figure}

It can be seen that the two most massive models are stable without the convective
correction.  This appears to be due to radiative damping in the shelf.  In
Figure~\ref{fig:eig938}, the first two nodes of the real part of the displacement occur
$r\approx 87\,\RSun$ and $141\,\RSun$, where the imaginary part is close to a local
maximum and minimum, respectively.  (Recall that the shelf begins at $46\RSun$.)  Thus the
phase increases with radius, as for an outward-propagating acoustic wave.  Evidently this
wave is damped almost completely, because if it were not, then upon reflection from the
photosphere a standing wave would result with real and imaginary parts in phase.  The
escaping acoustic power can be estimated as
$\dot E_{\rm ac} = 2\pi r^2\rho c_{\rm s} |\delta v|^2$, where
$\delta v = -i\omega\delta r$ is the radial velocity perturbation and
$c_{\rm s}\equiv(\Gamma_1P/\rho)^{1/2}$ is the adiabatic sound speed.  Evaluating this at
the first node and dividing by twice the total mode energy,
$2E_{\rm mode}= 4\pi\int \rho r^2 |\delta v|^2\dd r$, yields an estimate for the damping
rate of the mode amplitude: $11\,\mathrm{Md}^{-1}$.  This agrees well with the directly
calculated growth rate shown for this model in the fifth column of
Table~\ref{tab:growth_rates}.  Of course the calculations are not independent because the
estimate above uses the nonadiabatic eigenfunction.  But it does suggest that the
nonadiabatic calculation is self-consistent, and also that acoustic radiation into the
shelf is the dominant loss mechanism at our highest masses.  As further evidence of this,
the final column of Table~\ref{tab:growth_rates} shows growth rates calculated when the
`photospheric' boundary conditions \eqref{eq:bcout1} and \eqref{eq:bcout2} are imposed at
the local minimum of the pressure scale height (which does not exist in the two
least-massive models), thus effectively discarding the shelf region from the linear
analysis.  This results in a small positive growth rate.

\subsection{Strange modes}
\label{subsec:strange}

\begin{figure}
  \centering
  \includegraphics[width=\columnwidth]{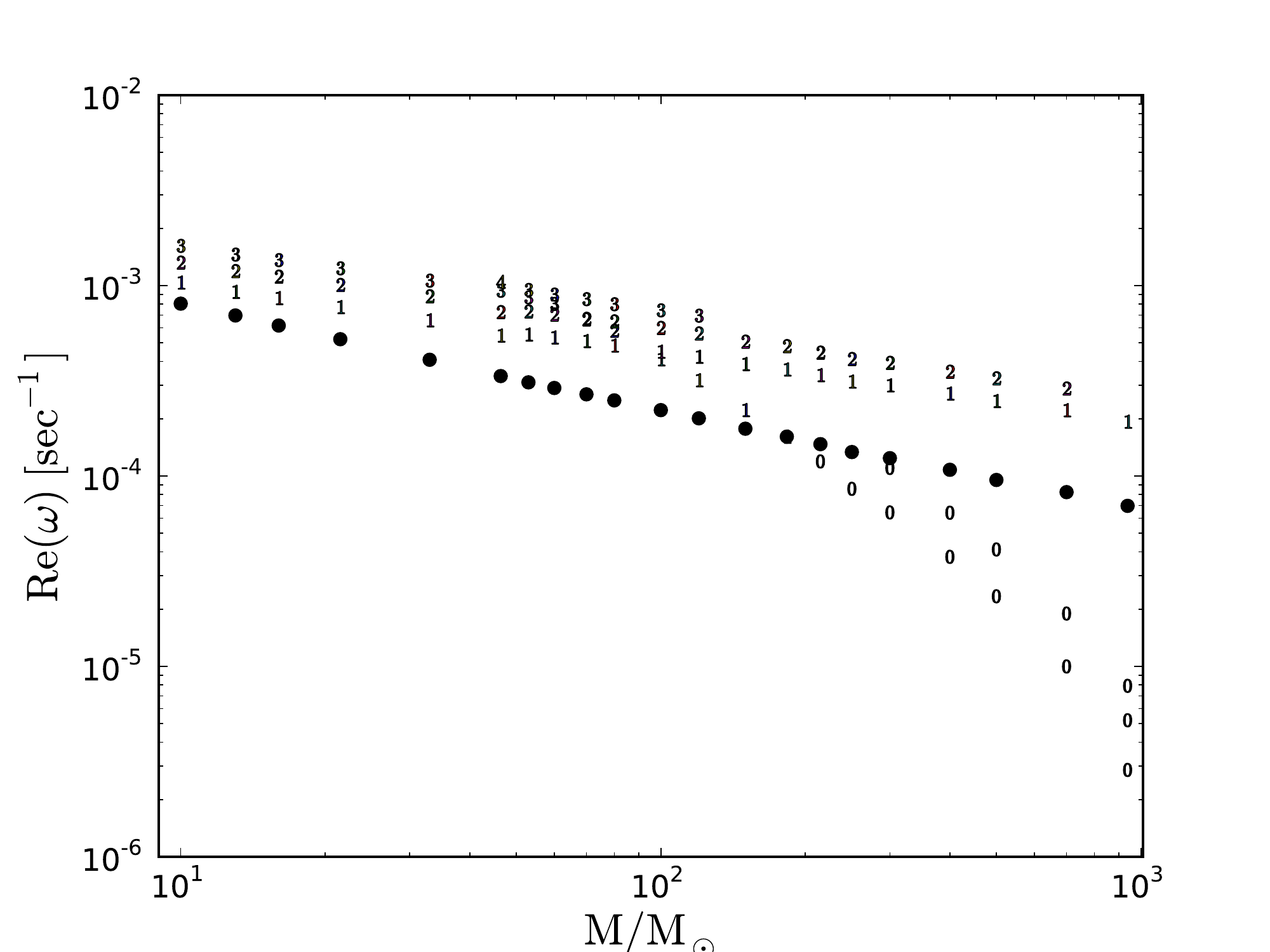}\\
  \includegraphics[width=\columnwidth]{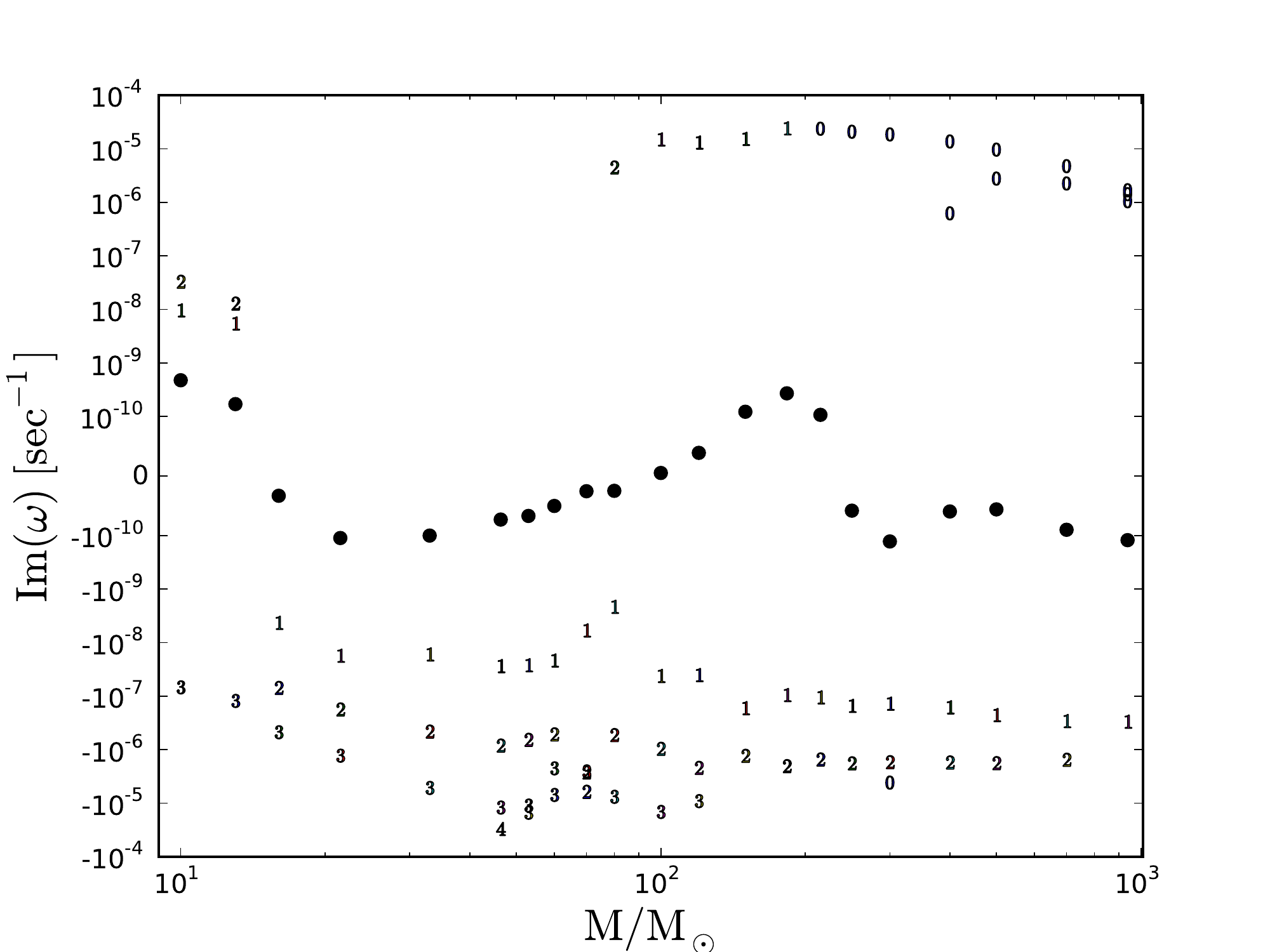}\\
  \caption{Real (first panel) and imaginary (second panel) parts of the lowest-lying modes
    versus stellar mass.  (Positive imaginary parts indicate instability.)  Solid circles
    mark the fundamental.  Other modes are marked by the number of
    nodes of $\delta r/r$ in the core.  $M/\MSun\in\{$10, 13, 16, 21.5, 33, 46.4, 53, 60,
  70, 80, 90, 100, 120, 150, 183, 200, 215, 250, 300, 400, 465, 500, 700, 983$\}$.
  \label{fig:bothparts}
}

\end{figure}

\begin{figure}
  \centering
  \includegraphics[width=\columnwidth]{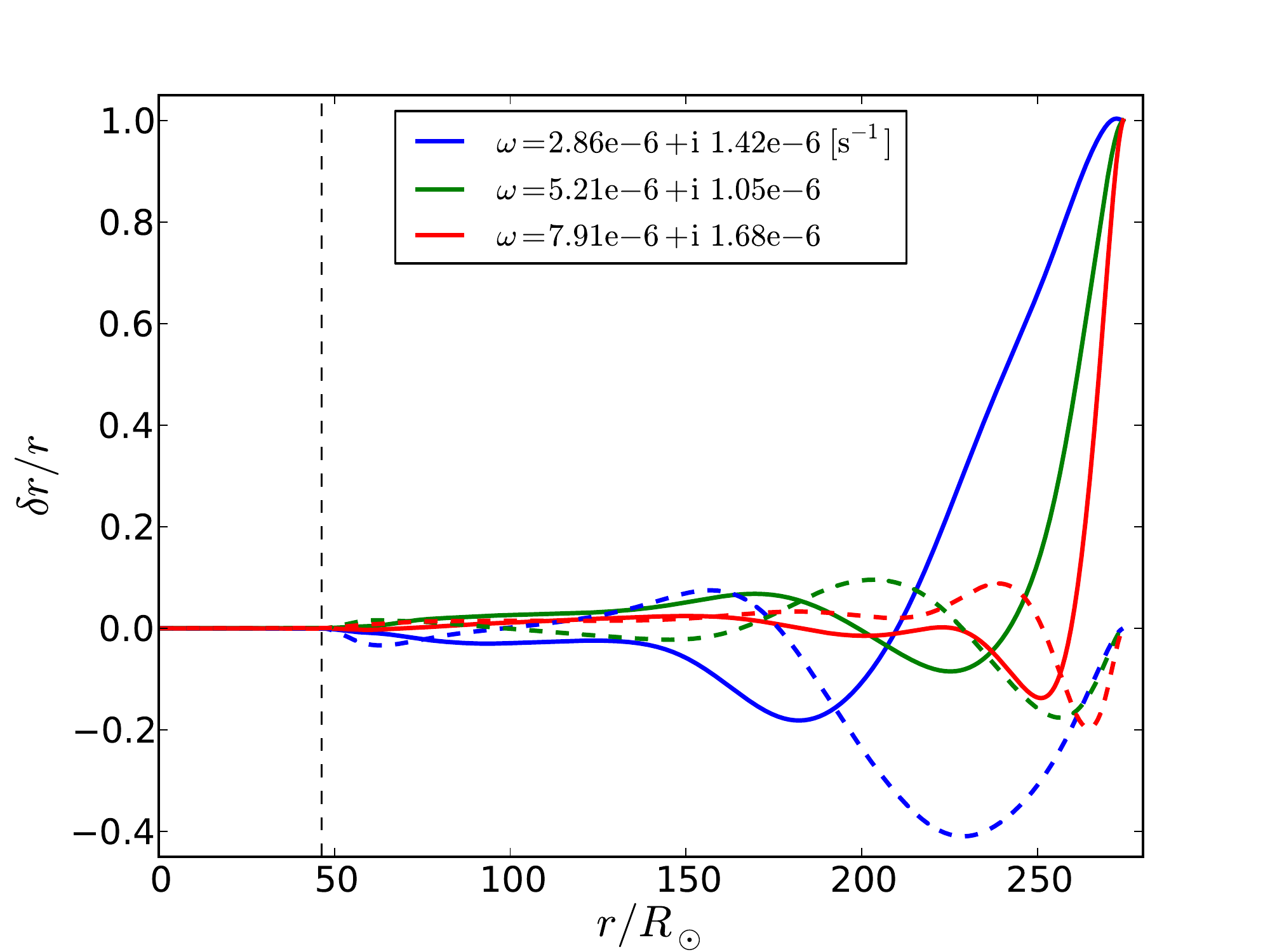}
  \caption{Three strange modes of the $938\MSun$ model, with frequencies as shown.  Solid
    and dashed curves show real and imaginary parts of the eigenfunctions, respectively.
    Thin vertical dashed line divides the core from the shelf (see
    Fig.~\ref{fig:shelf}). Note eigenfunctions are not weighted by $r^2\rho^{1/2}$.}
  \label{fig:strangers}
\end{figure}

In addition to the fundamental mode, there are in principle an infinite number of other
radial modes.  Some, like the fundamental itself, are slight modifications of adiabatic
counterparts and are concentrated in the core (the region below the local minimum of the
pressure scale height).  The real parts of the frequencies of these increase with the
number of nodes in the displacement eigenfunction ($\delta r/r$), while the imaginary
parts tend to become more negative (i.e., more damped) but are generally small because of
the long thermal time in the core.

Besides these, there are modes that have no obvious adiabatic counterpart, and which are
probably the strange modes discussed by \cite{Glatzel+Kiriakidis1993} and
\cite{Papaloizou+etal1997}.  Though some of these are damped, others have positive growth
rates [$\mbox{Im}(\omega)>0$] that approach the dynamical time.  Their energies are
strongly concentrated in the tenuous shelf region and thus are not likely to affect
directly the bulk of the star.  This is discussed more quantitatively below (\S\ref{subsec:strange_sat}).

Figure~\ref{fig:bothparts} shows the complex eigenfrequencies versus mass for the models in
Table~\ref{tab:models} and Fig.~\ref{fig:bothparts}.  For each mass, we show the four or five modes
with smallest real part---at least among those we have found.\footnote{It is possible that
  some low-lying modes have been missed.  At the higher masses, the search for modes
  becomes tedious, especially for the strongly nonadiabatic modes.  There are several
  causes, but the most pernicious is that each complex zero of the objective function
  described in the Appendix is paired with a pole, and the separation between poles and
  zeros decreases with increasing mass. This makes it difficult for zero-finders to
  ``smell'' their quarry from a distance in the complex plane.}  Below approximately
$50\MSun$, these are of the nearly adiabatic variety.  At about $53\MSun$, however, the
fourth and fifth harmonics cross: the real parts of their frequencies become nearly equal,
and their eigenfunctions differ mainly in the nascent shelf region.  By $60\MSun$, one of
the pair---the one with smaller real part---has its energy almost wholly concentrated in
the shelf.  By $70\MSun$, the successor to this mode crosses the second harmonic; by
$94\MSun$ it has crossed the first harmonic; and it crosses the fundamental at $183\MSun$.
Up to about $70\MSun$, the strange modes are damped, but at $M\ge 80\MSun$ they are grow
vigorously (Fig.~\ref{fig:bothparts}).  Our most massive model has at least three strange
modes with real parts of their eigenfrequencies below that of the fundamental, and growth
times on the order of ten days (Fig.~\ref{fig:strangers}).  All but $\sim10^{-8}$ of the
energies of these three modes lie in the shelf.

Our identification of the rapidly growing shelf modes with the strange modes is based
largely on the variation of their eigenfrequencies with mass and the concentration of
their eigenfunctions in the shelf.  \cite{Papaloizou+etal1997} propose as the definitive
test of ``strangeness'' that the modes should persist in the limit that
$\omega t_{\rm thermal}\to 0$.  We have not made this test but note that the shelf regions
of our most massive models are even more extreme than those of
\cite{Glatzel+Kiriakidis1993} with respect to radiation-pressure dominance and short
thermal times.
We refer the reader to \cite{Papaloizou+etal1997} and \cite{Glatzel+Kiriakidis1993} for
further discussion of the linear physics of strange modes.

\section{Nonlinear Saturation}
\label{sec:nonlinear}

\citeauthor{Appenzeller1970} proposed that radial pulsations of very massive stars saturate in
shocks that eject mass. His criterion for the onset of shocks was that the radial velocity at the
photosphere become larger than the local sound speed. \citeauthor{Papaloizou1973a} found in his own
numerical calculations that shocks were not so easily formed, and saturation occurred without mass
loss. Our view is closer to Papaloizou's, but we emphasize coupling to nonradial modes rather than
radial overtones, at least for the fundamental and other near-adiabatic modes.  
Saturation of strange modes is more likely to yield shocks and is discussed briefly in
\S\ref{subsec:strange_sat} below.

The largest growth rates we find are $\lesssim 10 \tKH^{-1}$. Here $\tKH$ is the Kelvin-Helmholtz
time defined as by \citet{Goodman+Tan2004}, which asymptotes to $\tKH \approx 3000\unit{yr}$ in the
limit of very large masses.\footnote{In order that the estimate of $\tKH$ not be biased by the
  extended but almost massless shelf, we use for $R$ the radius of the base of the shelf as defined
  in \S\ref{sec:atmosphere}.}  On the other hand, the pulsation periods recorded in
Table~\ref{tab:growth_rates} are $\lesssim 1\unit{d}$, and we expect this to scale $\propto
M^{-1/2}$ at higher masses. Thus the growth times are on the order of $10^5$ pulsation periods. In
this sense, the instabilities are extremely weak, even if the possibly stabilizing influence of
convection is ignored.

At one level, this is not a surprise. Whether caused by the $\epsilon$- or $\kappa$-mechanism,
pulsational instability operates by modulating the heat content of the star on the pulsation period.
Thus, the smallness of the ratio $\lvert\omegai/\omegar\rvert$ reflects the disparity between the
characteristic thermal and dynamical times of the star. We shall shortly argue that the smallness of
the linear growth rate implies a small amplitude at nonlinear saturation. 

This is not inconsistent with the relatively large amplitudes of oscillation of classical Cepheids
($\delta R/R \sim 0.1$) because the linear growth times are only $\sim 100$ pulsation periods in
those stars (e.g., \citealt{Castor1971,Bono+etal1999}), and it is worth recalling why
\citep[e.g.][]{Cox+Giuli1968}. Classical Cepheids are evolved stars with degenerate cores and a very
large ratio of central to mean density. Consequently, the eigenfunction $\zeta(r) \equiv \delta r/r$
of the fundamental radial mode is very far from homologous, $\zeta(0)/\zeta(R) \sim
\bar{\rho}/\rho(0) \ll 1$. The mode mass -- the factor by which one multiplies the mean-square
radial velocity at the surface to get the total energy in the mode -- is many orders of magnitude
smaller than the total mass of the star. In other words, for a given surface amplitude $\delta
R/R$, the stored energy in the mode is much less than it would be if the pulsations were homologous,
by a factor $\propto \Mmode/M_*$. Since the driving regions for the $\kappa$-mechanism lie near the
surface, the work integral is insensitive to the mode mass: it is of order $\Pi_0\,\delta L\,\delta
R/R$, where $\delta L/L \sim \delta R/R\sim -\delta T/T$ is the modulation of the surface luminosity
and $\Pi_0 \equiv 2\pi\omegar^{-1}$ is the pulsation period. Therefore the growth rate, which scales
with the ratio of the work done per cycle to the stored energy in the mode, is $\sim
(M_*/\Mmode)\tKH^{-1}$. The $\epsilon$ mechanism is negligible in Cepheids because the
nuclear-burning regions are in shell sources near the centre, where $\delta\log T$ and $\delta r/r$
are much smaller than in the ionization zones.

As a quantitative example, we have used \Mesa{} to create a `Cepheid' with the following parameters:
$M_* = 5.7~\MSun$, $R_* = 28.85~\RSun$, $\Teff = 5900\unit{K}$, and $L = 906~\LSun$. For this model,
$\bar{\rho}/\rho(0) = 6.2\times10^{-8}$, while $\zeta(0)/\zeta(R_*) = 3.3\times10^{-6}$ and
$\Mmode/M_* = 8.4\times10^{-5}$ for the fundamental radial mode computed with \Adipls. By contrast,
for the $938~\MSun$ main-sequence model we find $\zeta(0)/\zeta(R) = 0.66$ and $\Mmode/M_* =
0.061$.\footnote{This depends upon what one considers to be the stellar `surface.'  For the purpose
  of calculating $\Mmode$, we use $R_{\rm shelf}$ (\S\ref{sec:atmosphere}) when this is distinctly
  less than the photospheric radius.}

Thus the fundamental mode is approximately homologous and involves a significant fraction of the
star's mass. We expect $\Mmode/M_*$ to be nearly constant and comparable to this for larger masses
because of the similarity of these models to isentropic $n = 3$, $\Gamma_1 = 4/3$ polytropes.
\emph{Thus we also expect the growth rates of the fundamental radial mode to remain $\lesssim
10\tKH^{-1}$,  due to the relatively homogeneous structure and homologous pulsation of
these very massive main-sequence models, regardless of the details of the excitation 
mechanisms and of the shelf or wind.}

\subsection{Saturation of the fundamental via $3$-mode coupling}
\label{sec:nonlinear:coupling}

As a general rule, instabilities with smaller linear growth rates saturate at lower amplitudes. A
simple model equation for the amplitude envelope $A > 0$ might be
\begin{equation}\label{eq:ampeqn}
  \frac{dA}{dt} = \omegai A - \nu A^{n+1}\text{,}
\end{equation}
in which $\omegai$ is the linear growth rate, while $\nu$ and $n$ describe the nonlinearities. If
$\omegai$, $\nu$, and $n$ are all positive, then equilibrium is reached at $A_\mathrm{sat} =
(\omegai/\nu)^{1/n}$. As noted above, pulsations in classical Cepheids grow relatively rapidly.
Saturation in these stars and in RR~Lyraes occurs via shocks and nonlinear modification of the
conditions in the driving (ionization) zones \citep{Christy1966}. Because of their much smaller
dimensionless growth rates ($\omegai/\omegar \sim 10^{-6}$ instead of $\sim 10^{-3}$), the massive
main-sequence stars considered here may saturate at lower amplitudes, where more delicate
nonlinearities may be effective.

\citet{Dziembowski1982} suggested that $3$-mode coupling is responsible for saturation in dwarf
Cepheids, and that this explains why they oscillate at lower amplitudes than do classical Cepheids.
He worked directly from nonlinear equations of motion.  But if dissipation is weak, an action
principle, which is necessarily adiabatic, can be efficient in describing mode couplings.  The
lagrangian density $\mathcal{L}(\bxi,\bxidot)$ is expanded in powers of the displacement $\bxi$ and
its time derivative $\bxidot$.  The quadratic terms ($\mathcal{L}_2$) yield the linearized equations
of motion, while cubic and higher terms ($\mathcal{L}_3+\mathcal{L}_4+\ldots$) describe nonlinear
couplings.  When the amplitude of the primary/`parent' mode grows slowly from small amplitudes,
the cubic nonlinearities are the first to come into play. The most important couplings are those
that are resonant, meaning that the linear eigenfrequencies of the parent and daughter modes satisfy
$\omegap \approx \omega_\mathrm{d1} + \omega_\mathrm{d2}$, so that secular transfers of energy can
occur.\footnote{In resonance conditions such as this, all frequencies are understood to be real and
  nonnegative.}  However, the corresponding hamiltonian density $\mathcal{H}_2+\mathcal{H}_3$ cannot
be positive definite since the components of $\bxi$ can have either sign, so that the higher-order
nonlinearities must dominate if the amplitudes pass some threshold, perhaps leading to shocks and a
breakdown of the lagrangian description.  Actually, even when the amplitudes remain small,
dissipative terms must be added to the equations of motion to describe the linear damping of the
daughter modes. In application to pulsating stars, the growth rate of the parent is also represented
by a non-adiabatic term. The daughter modes are usually smaller in wavelength than the parent and
therefore more easily damped by radiative diffusion or (perhaps) eddy viscosity.

Landmark applications of $3$-mode coupling to the saturation of stellar instabilities include those
by \citet{Wu+Goldreich2001} (to white dwarf/ZZ~Ceti stars), as well as \citet{Schenk+etal2002} and
\citet{Arras+etal2003} (to rapidly rotating neutron stars). \citet{Papaloizou1973a} argued that
pulsations of very massive stars driven by the $\epsilon$-mechanism can saturate via direct resonant
couplings: that is, the coupling of a quadratic or higher power of the fundamental mode to a
\emph{higher}-frequency radial p~mode (overtone), so that $n\omegaf \approx \omegad$ for some
integer $n > 1$. Nonradial daughter modes offer many more possibilities for resonance, however:
g~modes are necessarily nonradial and have low frequencies, which is important for resonances of the
type $\omegap \approx \omega_\mathrm{d1} + \omega_\mathrm{d2}$ because the frequency of the
fundamental, $\omegaf$, is somewhat lower relative to the characteristic dynamical frequency
$\omega_* \equiv (GM_*/R_*^3)^{1/2}$ than in less massive stars, the ratio $\omegaf/\omega_*$
scaling as $M^{-1/2}$ \citep{Goodman+Tan2004}. Therefore we focus on couplings of this type. When
the parent is the radial fundamental mode, the strongest $3$-mode couplings are usually parametric
subharmonic, meaning that the two daughter modes are two copies of the same mode, with frequency
$\omegad \approx \omegaf/2$.  The eigenfunction of a typical daughter mode is high-order, with many
nodes in radius and angle, but its square is nonnegative and hence may have a significant $3$-mode
coupling with the nodeless radial fundamental.  Parametric subharmonic destabilization of g~modes
and internal waves has has been studied experimentally as well as theoretically \citep[and
references therein]{Benielli+Sommeria1998}.

We use our most massive ($938~\MSun$) model as an example.
Most of the star convects, but there is a radiative zone at $29 \lesssim r/\RSun \le 41$
containing $0.026M$. (There is also a second radiative zone at $r \ge 44~\RSun$, and still others in
the shelf, but these contain much less mass, so we neglect them here.)
The peak of the \BV{} frequency is $N_\mathrm{max} = 2.94\omega_*$, whereas the frequency of the
fundamental radial mode is $\omegaf = 1.1446\omega_* \approx 7.116\times10^{-5}\unit{rad~s^{-1}}$
according to \Adipls{} (which computes only the real part), in good agreement with our non-adiabatic
code. Thus there are many g~modes with frequencies $\sim \omegaf/2$. Using the approximate WKB
dispersion for high-order g~modes,
\begin{equation} \label{eq:WKB}
  \frac{\sqrt{l(l+1)}}{\omega_{ln}} \int N(r) \frac{\dr}{r} \approx \pi (n-\tfrac{1}{2})
\end{equation}
(where $n\ge 1$ counts radial nodes)
and the profile of the \BV{} frequency in the radiative zone, $N(r)$, we estimate that $n/l \approx
0.34$ for $\omega_{ln} \approx \omegaf/2$ and $l,n \gg 1$. For a non-rotating spherical star, so
that the eigenfrequency is independent of spherical-harmonic order $m$, the number of mode
frequencies in a given interval $\Delta\omega$ near $\omegaf/2$ that correspond to g~modes of degree
$l' \le l$ scales as $0.17\,l^2 \Delta\log\omega$ when $l \gg 1$. Inverting this, the minimum $l$ at
which one expects to find modes in the interval $\Delta\omega$ is
\begin{equation} \label{eq:lmin}
  l_\mathrm{min}(\Delta\omega) \approx 2.4\left(\frac{\Delta\omega}{\omega}\right)^{-1/2}\text{.}
\end{equation}

The distance $\Delta\omega$ from exact subharmonic resonance at which daughter modes can grow
depends upon the linear damping rate of these modes, the amplitude of the parent, and the $3$-mode
coupling coefficient. For the damping time of high-order g~modes by radiative diffusion, we apply
\extref{4.8} of \citet{Dziembowski1982} to our $938~\MSun$ model:
\begin{equation} \label{eq:tdamp}
  t_\mathrm{damp} = \gamma_\dd^{-1} \approx 1.0 \left(\frac{30}{l}\,\frac{\omegaf}{2\omega}\right)^2\unit{yr}\text{.}
\end{equation}

We evaluate the $3$-mode coupling coefficient from \extref{A8} of \citet{Kumar+Goldreich1989}.
Their formula assumes that the adiabatic exponent $\Gamma_1$ is constant and neglects the Eulerian
perturbation to the gravitational potential, which, although essential for the eigenfrequency of the
fundamental mode, is unimportant for the coupling since most of the stellar mass lies interior to
the propagation region of the g~modes in our case. We make the further approximation that the
fundamental mode is exactly homologous, meaning that its radial displacement is given by $\delta
r(r,t) = \zeta r\cos(\omegaf t)$, with $\zeta$ spatially constant but growing slowly as
$\exp(\omegai t)$.  We approximate the daughter-mode eigenfunctions using WKB, neglecting terms of
relative order $(k_r H_P)^{-1}$, where $k_r \approx \sqrt{l(l+1)}N/\omega r$ is the radial
wavenumber and $H_P$ the pressure scale height, and find that
\begin{equation} \label{eq:coupling}
  \int \mathcal{H}_3 d^3\mathbf{x} = \ldots + \left(\frac{3\Gamma_1-1}{2}\right) 
\zeta\cos\omegaf t \int N^2\delta r_\dd^2 \,\dd M\text{.}
\end{equation}
Here `$\ldots$' represents all $3$-mode coupling other than the one of interest, while $\delta r_\dd
= q_\dd(t) \xi_{r,d}(r) Y_{lm}(\theta, \phi)$ is the radial displacement of the daughter mode, with
time-dependent amplitude $q_\dd(t)$.  For simplicity, we set $\Gamma_1 = 4/3$, though the
mass-averaged value of $\Gamma_1$ in the $938~\MSun$ model is $\approx1.37$.

The integrand of \eqref{eq:coupling} is twice the potential energy per unit mass of the g~mode.  The
coupling term can therefore be treated as though it were a time-dependent correction to the linear
dynamics of the daughter mode, whose amplitude evolves according to
\begin{equation} \label{eq:Mathieu}
  \ddot q_\dd + 2\gamma_\dd\dot{q}_\dd + \omegad^2 \left[1 + 3\zeta \cos(\omegaf t)\right]q_\dd = 0\text{.}
\end{equation}
As usual with such Mathieu equations, if $\zeta$ and $\gamma_\dd/\omegad$ are both small, then the
solutions for $q_\dd(t)$ are approximately sinusoidal but with envelopes varying as $\exp(st)$:
\begin{equation} \label{eq:spar}
  s = -\gamma_\dd \pm \frac{3\zeta\omegad}{4} \sqrt{1-\left(\frac{4\Delta\omegad}{3\zeta\omegad}\right)^2}\text{,} \quad \Delta\omegad \equiv \omegad - \frac{1}{2}\omegaf\text{.}
\end{equation}
Thus in order that the daughter mode should grow even at exact resonance ($\Delta\omegad = 0$), we
must have $\zeta > 4\gamma_\dd/3\omegad$.  Since $\gamma_\dd \propto l^2$ [eq.~\eqref{eq:tdamp}],
this sets an upper bound to the degrees of daughter modes that can be destabilized when the
amplitude of the fundamental is $\delta R/R = \zeta\ll 1$.  Yet $l$ must be large enough so that it
is probable to find eigenfrequencies within the range $\lvert\Delta\omegad\rvert \le
3\zeta\omegad/4$ for which the square root in \eqref{eq:Mathieu} is real. Thus in effect we must
evaluate $\gamma_\dd$ at the degree $l_\mathrm{min}$ that is found by setting $\Delta\log\omega
\approx \zeta$ in \eqref{eq:lmin}. Then $\gamma_\dd \propto l_\mathrm{min}^2 \propto \zeta^{-1}$, so
that the requirement $\tfrac{3}{4}\zeta\omegad > \gamma_\dd$ for growth leads to an inequality of
the form $\zeta > C\zeta^{-1}$. Evaluating the numerical factor $C$, we find that the threshold for
exciting daughter modes is approximately
\begin{equation} \label{eq:threshold}
  \left(\frac{\delta R}{R}\right)_\mathrm{min} \approx 3\times10^{-3}\text{,} \quad l_\dd \approx 45\text{.}
\end{equation}
Since the mode frequencies are sensitive to details of the stellar model, the occurrence of
resonance is effectively probabilistic, and therefore the threshold for subharmonic instability will
vary somewhat. In fact, for the particular model considered here, \Adipls{} finds
$\Delta\omegad/\omegad \approx 4\times10^{-4}$ at $l_\dd = 26$, which is some $20$ times closer to
resonance than would be expected from the statistical estimate \eqref{eq:lmin}.

The threshold \eqref{eq:threshold} of subharmonic instability involves only the current amplitude of
the fundamental mode, not its rate of growth, $\omegai$. The latter is important for deciding
whether the daughter modes can actually accept and dissipate energy from the parent faster than the
nonadiabatic work integral increases that energy. \citet{Arras+etal2003} state as a rule of thumb
that the condition for this is simply $\gamma_\dd > \omegai$, a regime they call `weak driving.'
Clearly this would suffice for nonlinear saturation of the unstable parent mode if a daughter mode
could reach energy equipartition with the parent, but that is unlikely in our case.  The wavelength
of the first daughters to go unstable is much smaller than the radius of the star, roughly by a
factor $2/l$ when one accounts for both the radial and angular components of the wavenumber. Also
the mass of the g-mode propagation zone is only $0.026 M$, whereas the effective mass of the
fundamental mode is $0.14 M$, as previously discussed. Therefore if a daughter mode at, say, $l_\dd
= 45$ were to have the same energy as the fundamental, it would have a strain rate (spatial
derivative of velocity) roughly $2\pi \times (l/2) \times \sqrt{0.14/0.026} \approx 7.\,l$ times
larger than the parent. At such a strain rate, the daughter mode would destabilize still other modes
(granddaughters) and probably transfer energy to them more quickly than it could receive energy from
the parent. Therefore equipartition is unlikely.

On the other hand, $\gamma_\dd \gg \omegai$ in the present case, so that the rate of linear
dissipation by daughter modes could balance the growth of the parent even if the daughters' energies
were well below equipartition with the parent. Because of the degeneracy of the eigenfrequencies in
a non-rotating star, $2l_\dd+1$ daughter modes grow at the same rate. When the average energy
\emph{per mode} reaches a value $\bar{E}_\dd$, the total linear dissipation rate becomes
$2(2l_\dd+1) \gamma_\dd \bar{E}_\dd$. Setting this equal to the rate at which the fundamental mode
gains energy from its own linear instability, $2\omegai E_f$, shows that saturation is possible when
$\bar{E}_\dd/E_f \approx (2l_\dd+1)^{-1}(\omegai/\gamma_\dd)$. Evaluating this for $l_\dd = 45$,
$\gamma_\dd^{-1} \approx 0.46\unit{yr}$ [cf.\ \eqref{eq:tdamp}], and $\omegai = (500\unit{yr})^{-1}$
leads to $\bar{E}_\dd/E_f \approx 10^{-5}$. The ratio of strain rates is then
\begin{equation}
  \frac{1}{2}\sqrt{\frac{\bar{E}_\dd}{E_f}} \times 4.6 l_\dd \approx 0.3 \quad (l_\dd \approx 45)
\end{equation}
(a factor of $1/2$ reflects the lower frequency of the daughters). Since this is less than unity,
saturation of the parent/fundamental mode is likely at daughter amplitudes too small to excite
granddaughters.

We conclude that it is indeed likely that $3$-mode coupling will saturate the growth of the
fundamental radial mode. However, some caveats are in order regarding rotation, which we have so far
neglected.

There are at least two rotational regimes to consider: slow and fast. Slow rotation at an angular
velocity $\Omega \gtrsim l_\dd^{-1}\omega_*$ but $\ll N_\mathrm{max}$ will lift the degeneracy with
respect to spherical-harmonic order $m$, while preserving the degree $l$ as a useful approximate
quantum number. Since there are more distinct eigenfrequencies, subharmonic resonance becomes
possible at smaller $l$: \eqref{eq:lmin} is replaced by $l_\mathrm{min} \approx 1.6
(\Delta\omega/\omega)^{-1/3} \to 0.6\zeta^{-1/3}$. Otherwise following the same steps as before, the
threshold of instability occurs at $\zeta \approx 5.4\times10^{-4}$ and $l_\dd \approx 20$ (both
lower than before). Now a single, nondegenerate daughter mode first goes unstable, so the required
balance at saturation if this daughter only is active becomes $2\gamma_\dd \bar{E}_\dd = 2\omegai
E_f$, and the ratio of strain rates (daughter:parent) works out to $\approx 3$ instead of $0.3$.
Hence nonlinear coupling of the daughter to granddaughters may occur, limiting the energy of the
former and complicating the analysis. On the other hand, the number of unstable daughters will
increase rapidly ($\propto\zeta^{5/2}$) as the amplitude of the parent increases above the first
subharmonic threshold, so without having analysed the situation carefully, we still expect
saturation to occur.

By fast rotation, we mean fast enough so that inertial oscillations -- approximately incompressible
motions restored by Coriolis rather than buoyancy forces -- can have resonant $3$-mode couplings
with the parent, as considered by \citet{Schenk+etal2002} and \citet{Arras+etal2003} for neutron
stars. Since the maximum frequency of inertial oscillations is $2\Omega$, a necessary condition for
subharmonic instability of the fundamental mode is $\Omega > \omegaf/4$. In the $938~\MSun$ model,
this translates to $\Omega > 0.286\omega_*$, which is half or less of the mass-shedding limit for
an $n = 3$ polytrope, depending how one defines $\omega_*$ for a nonspherical body
\citep{Hurley+Roberts1964}. Rapid rotation is not unreasonable for a body recently formed by
fragmentation of an AGN accretion disk. Furthermore, $\omegaf/\omega_*$ scales as $M^{-1/2}$ with
increasing stellar mass.\footnote{Uniform rotation at the mass-shedding limit may set a lower limit
  to $\omegaf/\omega_*$ because rotational energy behaves somewhat like gas pressure in the
  time-dependent virial theorem. Due to the central concentration of $n = 3$ polytropes, however, we
  estimate that this limit comes into effect only for $M \gtrsim 10^5~\MSun$, where relativistic
  corrections must also be considered \citep{Baumgarte+Shapiro1999}.} Unlike g~modes, inertial
oscillations propagate in convection zones, so that they may be destabilized throughout these
(largely convective) massive stars. This is salient because we are not sure how the mass fraction of
the radiative zones should scale at masses above $10^3~\MSun$ and super-solar metallicities.  Even
when the condition $\Omega > \omegaf/4$ is not satisfied, the imposition of rotation on vigorous
convection will surely lead to magnetic fields, perhaps in rough equipartition with the convection,
so that radial pulsations may couple nonlinearly to small-scale Alfv\'enic modes.

\subsection{Saturation of strange modes.}
\label{subsec:strange_sat}

Because of their large growth rates, strange modes are unlikely to saturate
via 3-mode couplings of the sort discussed above.  They will grow to large
amplitudes and probably saturate via shocks.

As a conservative criterion for the amplitude at which a shock appears, we take
$|\partial\delta r/\partial r|_{\rm\max}\ge 1$, since this would predict shell crossing in
the absence of shocks.  For the fastest-growing strange mode of the $938\MSun$ model,
$|\partial\delta r/\partial r|_{\rm\max} = 25.9 \delta R/R$.  The maximum is achieved at
$r = 0.984 R$.  Shocks may then appear when the surface amplitude
$\delta R/R= (25.9)^{-1}\approx 4\%$.  At this point, the amplitude of the displacement
eigenfunction at the surface of the core ($r\approx 0.169 R$) will be only
$1.4\times 10^{-6}$.  The corresponding numbers for the other two strange modes of this
model are $2\times10^{-6}$ and $7\times10^{-6}$.  Given the smallness of these numbers, it
seems inconceivable that strange modes could threaten the survival of the star on
dynamical timescales.

Whether finite-amplitude strange modes drive mass loss from the shelf is a different and
difficult question.  On the one hand, the maximum radial velocity at shock onset is rather
small ($\lesssim5\%$) compared to the escape velocity $v_{\rm esc}=(2GM/R)^{1/2}$.  On the
other hand, the residual between gravitational and radiative accelerations is relatively
small, so that escape may be possible at $v\ll v_{\rm esc}$.  Furthermore, line-driven
steady winds are likely even without the assistance of shocks.  If the mass-loss rate of
such a wind is high enough, it may tend to suppress the linear instability of the strange
modes.  All of this we leave for later investigation.

\section{Summary and Discussion}
\label{sec:conclusion}

We have re-examined the stability of the fundamental radial mode of very massive main-sequence
stars. Although nonradial and higher-order radial modes may also be unstable, we focus on the radial
fundamental because collapse or explosion of these radiation-pressure-dominated objects would begin
with this mode at linear order. In agreement with \citeauthor{Shiode+Quataert+Arras2012}, we find
that the linear growth rate is sensitive to turbulent convective damping. We have extended their
results to higher masses at solar metallicity, and we have used a fully nonadiabatic rather than
quasi-adiabatic method, which allows us to treat the $\kappa$-mechanism more reliably. The
$\epsilon$ mechanism is more important for our most massive models we consider, however.

The linear growth rates remain uncertain not only because of the turbulent bulk viscosity, but also
because of the tenuous (and possibly unphysical) envelopes possessed by all of our models above
$100~\MSun$. In fact we find negative growth rates even without convection, apparently due to
radiative damping in the shelf.  Nevertheless, the growth rate should in any case be extremely small
even if positive, $\omegai/\omegar \sim \Pi_0/\tKH$, due to the relatively low central
concentrations of these stars and correspondingly large mode masses.

We have then argued from the smallness of the linear growth rate (in case this is positive) that the
radial fundamental should saturate at a small amplitude due to any one of a number of weak
nonlinearities. To support this claim, we have estimated the saturation amplitude that would result
if parametric coupling to high-order g~modes were the most important nonlinearity. For our most
massive model, the estimate is $\delta R/R \approx 3\times10^{-3}$. Other nonlinear couplings may
stop the growth at even smaller amplitudes, but those we identify would depend on uncertain
parameters such as the star's rotation rate or magnetic field.

We have also shown that our models, like those of GK93, are subject to a class
of intrinsically nonadiabatic modes having much larger growth rates but confined to the
shelf: strange modes.  These we estimate to saturate at fractional surface displacements
of a few percent via shocks. Their contribution to mass loss, if any, can be reliably
estimated only in the context of a time-dependent wind model that includes a number of
other nonlinear effects, such as line driving.  However, even at saturation, the energy
of the strange modes in the stellar core is neglegible, and therefore they probably
affect the bulk of the star only secularly.

A number of physical simplifications and compromises have been made:
restriction to solar metallicity; neglect of rotation; and neglect of perturbations to
the convective flux.  Increased metallicity might produce even more extended
``shelves'' in the equilibrium models, and larger growth rates for the modes driven by
the epsilon mechanism. However, presuming that the growth rates of the fundamental
mode varied roughly linearly with $Z$, they would remain very small compared to the
dynamical time even at metallicities ten times solar, such as may obtain in AGN disks
\citep{Dietrich+etal2003,Nagao+etal2006}.  Rotation is expected to have a
stabilizing influence on the fundamental mode at very high masses because it
contributes to the perturbed energy under homologous changes in radius somewhat like
gas rather than radiation pressure \citep{Baumgarte+Shapiro1999}. This could be
important for very massive stars formed in an AGN disk, and perhaps continually
accreting from that disk, since such objects would probably rotate rapidly
\citep{Goodman+Tan2004,Jiang+Goodman2011}.  Neglect of perturbations to the convective
flux has surely caused quantitative errors in the growth rates.  \cite{Guzik+Lovekin2012},
using a prescription for such perturbations that incorporates a time delay in the
convective response, find that super-Eddington luminosities can occur during part of the
pulsation cycle, perhaps leading to mass loss.  However, their analysis is limited to the
outer parts of the star.  More importantly, their prescription relies on mixing-length
theory, which may not be reliable in the extremely-radiation-pressure-dominated shelf
regions \citep{Jiang+etal2015}.

Despite these simplifications and uncertainties, it seems likely that the
growth rate of the fundamental mode must be extremely small compared to the reciprocal
of the dynamical time, and therefore that pulsations in the
fundamental will saturate nonlinearly at small amplitudes too small to disrupt or collapse
the star---at least on the main sequence.
We conclude that thermally driven pulsations of the radial fundamental mode do not limit the
main-sequence lifetimes of very massive stars. The tenuous outer envelopes of the more massive
\Mesa{} models, however, which stem from an opacity bump at $\sim 10^5\unit{K}$, lead us to suspect
that these stars would have powerful winds if the hydrostatic constraint were lifted, and that the
mass-loss timescale ($M_*/\dot{M}$) may be much less than one million years, though necessarily
longer than the Kelvin-Helmholtz timescale ($\approx 3000\unit{yr}$). The lower bound would be
achieved only if all of the stellar luminosity were converted to the mechanical energy of a wind
with vanishing asymptotic velocity at infinity.  For a very massive star embedded in a dense AGN
disk, continued accretion from the disk might easily exceed the wind losses, perhaps causing it to
grow to such a mass as to undergo relativistic instability.

These results suggest a few directions for future research.  It will be
relatively straightforward to explore the effect of super-solar metallicities on the
linear growth rates.  Changes in the growth rates as the models evolve away from the
zero-age main sequence could also be studied, although we have not yet succeeded in
evolving our most massive \Mesa{} models to the end of their main-sequence phases.
Probably more important, but also more challenging, will be to determine the mass-loss
timescale.  Several physical mechanisms will have to be considered, including
line-driven winds \citep{CAK75}; inhomogeneous optically-thick winds \citep{Owocki2015};
and perhaps winds driven by nonlinear strange modes or other radiation-driven
instabilities.  Still more mechanisms may operate in late stages of stellar evolution,
such as wave-driven winds \citep{Quataert+etal2015}.  The range of possibilities is
narrowed if one focuses on mechanisms that operate early in the life of a star and that
are capable of removing much of its initial mass in much less than the nominal
main-sequence lifetime.  Even so, multi-dimensional calculations with
frequency-dependent radiative transfer may be required.

\section*{Acknowledgements}

This project made extensive use of the \Mesa{} stellar-evolution code and the \Adipls{}
asteroseismology package.

\appendix
\section{Method for stiff linear boundary-value problems}
\label{sec:stiff}

The method described here is similar  to that of \citet{Drury1980} and \citet{Davey1983},
but different in detail and slightly simpler, at least in derivation.
One wants to solve
\begin{equation}
  \label{eq:dydx}
  \frac{\dd\by}{\dd x} = \bA\by\,
\end{equation}
$\by$ being a column vector of length $n$ representing the dependent variables, and $\bA$ an
$n\times n$ matrix depending upon an eigenvalue to be determined, and usually also on the
independent variable, $x$.  In our case $n=4$, and the eigenvalue is the complex frequency of pulsation,
$\omega$.  There are $p$ homogeneous boundary conditions to be satisfied at the left
boundary, $x=x_{\min}$, and $n-p$ at the right boundary, $x=x_{\rm max}$.  These are represented
by $p\times n$ and $(n-p)\times n$ matrices $\bB(x_{\min})$ and $\bC(x_{\max})$:
\begin{equation}
  \label{eq:bcsy}
  \bB(x_{\min})\by(x_{\min}) = 0\,,\qquad \bC(x_{\max})\by(x_{\max}) = 0\,.
\end{equation}
One expects nonzero solutions for $\by(x)$ only for discrete values of $\omega$, which are to be
determined.

\begin{figure}
  \centering
  \includegraphics[width=\columnwidth]{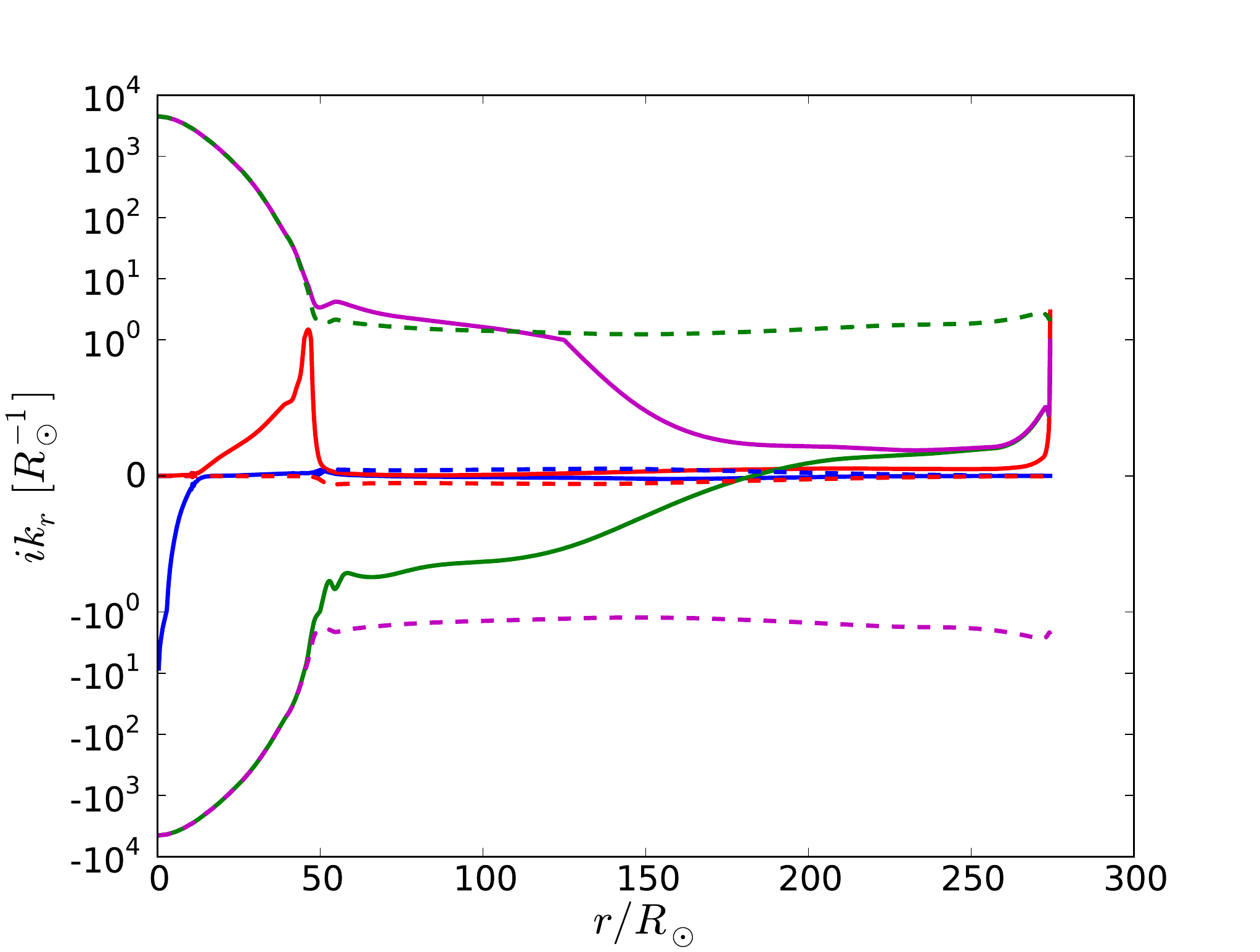}
  \caption{Real (solid) and imaginary (dashed) parts of the 4 eigenvalues of $\bA$ in the
    $938\,M_\odot$ model. Vertical scale is linear within the interval
    $(-R_\odot^{-1},R_\odot^{-1})$, and logarithmic outside it.}
  \label{fig:krs}
\end{figure}

The difficulty in solving this boundary-value problem is that the $n$ complex eigenvalues of
$\bA(x,\omega)$ may differ widely in size (Fig.~\ref{fig:krs}).  In our problem, they vary over 4-6
orders of magnitude in the inner parts of the star, due to the large ratio of thermal to dynamical
time.\footnote{ The eigenvalues $\{\lambda_1,\ldots,\lambda_n\}$ of $\bA$ at a particular $x$ should
  not be confused with the eigenvalue $\omega$ of the entire boundary-value problem.  It might be
  better to speak of wavenumbers $k_x\equiv-i\lambda$.  Since $\bA$ depends upon $\omega$ as well as
  $x$, the $\lambda$s and $k_x$s do as well.}  Furthermore, since the eigenfunction corresponding to
$\lambda$ behaves as $\sim\exp(\int^x\lambda dx)$, direct integration of eq.~\eqref{eq:dydx} can
overflow or underflow machine precision if the real part of $\lambda$ is large.  This happens in our
stellar problem, and both signs of $\mbox{Real}(\lambda)$ occur simultaneously.  It therefore proves
impractical to use a conventional shooting method in which one iteratively makes guesses for the
unconstrained components of $\by$ at both boundaries (and for $\omega$) and integrates toward a
fitting point.  As described above, a relaxation (Henyey-type) method also failed.

Instead, we propagate the boundary conditions themselves to the fitting point.  For a given choice
of $\omega$ and for any $x$ and $x'$ between the boundaries, eq.~\eqref{eq:dydx} has the formal
solution $\by(x) =\bP(x,x')\by(x')$ if the `propagator' $\bP$ satisfies
\begin{multline}
  \label{eq:Pprop}
\frac{\partial\bP}{\partial x}(x,x')=\bA(x)\bP(x,x'),\quad
\frac{\partial\bP}{\partial x'}(x,x')= -\bP(x,x')\bA(x'),\\
\bP(x,x')=\bP(x,x'')\bP(x'',x'),\quad   \bP(x,x) = \mathbf{I},
\end{multline}
$\mathbf{I}$ being the $n\times n$ identity.  The lefthand boundary
condition \eqref{eq:bcsy} can then be restated as $\bB(\bar x)\by(\bar x)=0$ with
$\bB(\bar x)\equiv\bB(x_{\min})\bP(x_{\min},\bar x)$.
Similarly $\bC(\bar x)\by(\bar x)=0$ with $\bC(\bar x)\equiv\bC(x_{\max})\bP(x_{\max},\bar x)$.
In other words, the solution $\by(\bar x)$ at any intermediate $\bar x$  between $x_{\min}$ and $x_{\max}$
must belong to the subspace annihilated by $\bB(\bar x)$, and also to the subspace annihilated by
$\bC(\bar x)$.  Since the dimensions of these two subspaces add up to $n$, their intersection is
only $\by=0$ unless $\omega$ is a root of
\begin{equation}
  \label{eq:detBC}
  \det\{\bB(\bar x),\bC(\bar x)\} = 0\,,
\end{equation}
$\{\bB,\bC\}$ meaning the $n\times n$ matrix whose first $p$ rows coincide with those of $\bB$ and
whose last $n-p$ rows those of $\bC$.

This reformulation may appear pointless since it is no easier to solve for $\bP$ by
direct integration of eq.~\eqref{eq:bcsy} than to solve eq.~\eqref{eq:dydx} itself.  The same
difficulties with stiffness and overflow occur.  Furthermore, as the separation between $\bar x$ and
$x_{\min}$ increases, the rows of $\bB(\bar x)$ are dominated by the fastest-growing
eigenvector of $\bA$, so that they quickly become linearly dependent when estimated in finite-precision
arithmetic.  A key observation, however, is that the constraint $\bB(x)\by(x)=0$ is equivalent to
$\mathbf{L}\bB(x)\by(x)=0$ for any nonsingular $p\times p$ matrix $\mathbf{L}$.  This can be
exploited to keep the rows of $\mathbf{LB}(x)$ linearly independent, in fact orthonormal.

In practice one integrates \eqref{eq:dydx} or \eqref{eq:bcsy} on a discrete grid
$x_{\min}=x_0 < x_1 < x_2 \ldots < x_N = x_{\max}$.
Let $x_m$ be the fitting point, $0<m<N$.  The values of $\bB(x_k)$ could be defined iteratively
according to
\begin{equation*}
  \bB(x_{k+1}) =   \bB(x_k)\bP(x_k,x_{k+1}),\quad \bB(x_0)=\bB(x_{\min}).
\end{equation*}
Instead of this, however, we solve
\begin{equation}
  \label{eq:Biter}
    \mathbf{\hat B}_{k+1} =   \mathbf{G}_{k+1}\mathbf{\hat B}_k\mathbf{\hat P}_{k,k+1},\quad 
\mathbf{\hat B}_0=\bB(x_{\min}).
\end{equation}
Here $\mathbf{\hat P}_{k,k+1}$ is a discrete approximation to $\bP(x_k,x_{k+1})$---a suitable choice
will be given presently---while $\mathbf{G}_{k+1}$ is chosen at each step to make the rows of
$\mathbf{\hat B}_{k+1}$ orthonormal.  For example, if $p=2$ and the rows of
$\mathbf{\hat B}_k\mathbf{\hat P}_{k,k+1}$ are $\mathbf{a}$ and $\mathbf{b}$, then
\begin{equation}
  \label{eq:Gdef}
  \mathbf{G}_{k+1} = \begin{pmatrix}1&0\\ 0&\gamma\end{pmatrix}
\begin{pmatrix}1&0\\ -\beta & 1\end{pmatrix}
\begin{pmatrix}\alpha & 0 \\ 0 & 1\end{pmatrix}\,
\end{equation}
with $\alpha\equiv|\mathbf{a}|^{-1}$, $\beta\equiv\alpha\mathbf{ba^\dag}$, and
$\gamma = |\mathbf{b}-\alpha\beta\mathbf{a}|^{-1}$.  This represents the Gram-Schmidt process, which
can be extended to any number of rows. It is clear that if $\mathbf{\hat
  P}_{k+1}=\bP(x_k,x_{k+1})$, then $\mathbf{\hat B}_m=\mathbf{G}_0\mathbf{G}_1\ldots\mathbf{G}_m\bB(x_m)$,
so that $\det\bB(x_m)=0$ if and only if $\det\mathbf{\hat B}_m=0$.

We now discuss a suitable approximation for $\bP(x_k,x_{k+1})$.  If $\bA$ were constant over the
interval $[x_k,x_{k+1}]$, then $\mathbf{\hat P}_{k,k+1}=\exp[(x_{k+1}-x_k)\bA]$ would be exact, and
otherwise if $\bA$ is evaluated at the midpoint $x_{k+1/2}\equiv(x_k+x_{k+1})/2$, then the
exponential approximation formally second order in $\Delta x_k\equiv x_{k+1}-x_k$.  However, in our
problem, the largest eigenvalue of $\bA_{k+1/2}$ can be so large that on a reasonable mesh (we
typically interpolate the \Mesa{} model with splines at $\sim 10^3$-$10^4$ uniformly spaced radii),
the matrix exponential can overflow, or at least cause substantial loss of precision.  We therefore
adopt the Crank-Nicholson approximation
\begin{equation}
  \label{eq:Phat}
  \mathbf{\hat P}_{k,k+1}= (\mathbf{I}+\tfrac{1}{2}\Delta x_k\bA_{k+1/2})
 (\mathbf{I}-\tfrac{1}{2}\Delta x_k\bA_{k+1/2})^{-1}
\end{equation}
Eigenvalues of $\bA_{k+1/2}$ with positive (negative) real part, corresponding to behaviours that
grow (decay) with increasing $x$, are mapped to eigenvalues of $\mathbf{\hat P}_{k,k+1}$ inside
(outside) the unit circle, presuming $x_k<x_{k+1}$; very large eigenvalues of $\Delta x_k
\bA_{k+1/2}$ are mapped to eigenvalues of $\mathbf{\hat P}_{k,k+1}$ close to $-1$.  It is still
necessary to use a mesh fine enough so that the behaviours that should decay with the iteration
\eqref{eq:Biter} do so quickly enough; the growing behaviours are controlled by the
orthonormalization.  In practice this is determined by varying the mesh resolution and monitoring
the effect on the estimated root of \eqref{eq:detBC} for $\omega$.  

Another constraint on the mesh is that the Crank-Nicholson approximation \eqref{eq:Phat} for
$\bP(x_k,x_{k+1})$ encounters a pole if $\bA_{k+1/2}$ has $2/\Delta x_{k+1/2}$ as an eigenvalue.  In
fact, the eigenvalue represented by the blue lines in Fig.~\ref{fig:krs} is nearly real and diverges
toward the centre of the star.  This corresponds approximately to the singular solution of the
adiabatic equation, $\propto r^{-3}$ as $r\to 0$.  To control this, we start the integration at a
nonzero but small radius $r_{\min}$ with a mesh spacing $\Delta r\ll r_{\min}/3$.  This ensures that
no poles are encountered as we integrate outward, since the other two large eigenvalues, which
represent strongly non-adiabatic heat diffusion, have almost equal real and imaginary
parts,\footnote{They satisfy $-i\omega\delta T \approx\eta\partial^2\delta T/\partial r^2$, where
  $\eta=16\sigma_{\textsc{sb}}T^3/3\kappa\rho^2 c_P$ is the thermal diffusivity.  Hence the
  eigenvalues $ik_r\approx\pm\sqrt{-i\omega/\eta}$.}  so that they cannot cause poles
for real values of $\Delta x_k=\Delta r$.

The righthand boundary condition is translated to the matching point by constructing the
$(n-p)\times p$ matrix $\mathbf{\hat C}_m$ in similar fashion.

\subsection{Analyticity}

In many problems of physical interest, the components of the matrix $\bA$ depend analytically or
even polynomially on the eigenvalue $\omega$.  Such is the case in our stellar problem, since
$\omega$ is a temporal frequency and there are a finite number of time derivatives in the linearized
equations.  The matrices $\bP(x,x')$, $\bB(x)$, $\bC(x)$, and the determinant \eqref{eq:detBC} can
then be expected also to depend analytically on $\omega$, i.e. their derivatives with respect to
$\omega$ satisfy the Cauchy-Riemann equations.  It is desirable to have a numerical
scheme that preserves this analyticity up to round-off error, notwithstanding truncation error.  For
one thing, if $F(\omega)$ represents the determinant \eqref{eq:detBC}, then the roots of
$F(\omega)=0$ can be sought by an efficient algorithm that assumes analyticity.  The simplest of
these, and what we use, is false position,
\begin{equation}
  \label{eq:secant}
  \hat\omega_{n+1} = \frac{\hat\omega_{n-1}\hat F(\hat\omega_n)-\hat\omega_{n}\hat
    F(\hat\omega_{n-1})}{\hat F(\hat\omega_n)-\hat F(\hat\omega_{n-1})}\,.
\end{equation}
This is only a minor convenience since the root could be sought by treating the
real and imaginary parts of $\hat\omega$ (and of $\hat F$) as independent.  More importantly, the
convergence of eq.~\eqref{eq:secant} for complex $\hat\omega$ depends upon the extent to which
\begin{equation}
  \label{eq:analyticity}
  \frac{\hat F(\omega + i\epsilon)-\hat F(\omega)}{i\epsilon}\approx
  \frac{\hat F(\omega + \epsilon)-\hat F(\omega)}{\epsilon}\,,\quad
|\epsilon|\ll |\omega|\,,
\end{equation}
and so can be used to monitor the effects of round-off error.

The numerical scheme described above is \emph{not} analytic in $\omega$, even in exact arithmetic,
because of the complex conjugations involved in the orthonormalization matrices $\mathbf{G}_k$ and
$\mathbf{L}$.  That is to say, if $\hat F(\omega)$ were defined as the determinant that results from
replacing $\bB(\bar x)$ and $\bC(\bar x)$ with $\mathbf{\hat B}_m$ and $\mathbf{\hat C}_m$ in
eq.~\eqref{eq:detBC}, then it would not be analytic in $\omega$.  However, analyticity can be
rescued by the following simple trick.  The condition $\mathbf{\hat B}_m\by_m=0$ constitutes $p$
constraints on the $n$ components of $\by_m$.  Therefore, it can be rewritten as
\begin{equation}
  \label{eq:vfromu}
\mathbf{v}=\mathbf{\tilde B}\mathbf{u},
\end{equation}
where $\mathbf{u}$ is a column vector containing the first $n-p$ components of $\by_m$, $\mathbf{v}$
contains the remaining $p$ components, and $\mathbf{\tilde B}$ is $p\times (n-p)$.  Now it is easy
to see that $\mathbf{\tilde B}$ is independent of the non-analytic orthonormalizing factors
$\mathbf{G}_0\ldots\mathbf{G}_N\equiv\mathbf{L}$ defined via eqs.~\eqref{eq:Biter}-\eqref{eq:Gdef},
at least in principle, since $\mathbf{\hat B}_m\by_m=0$ is equivalent to
$\mathbf{L}^{-1}\mathbf{\hat B}_m\by_m=0$, and the latter is analytic. Similarly, we can rewrite the
translated righthand boundary condition $\mathbf{\hat C}_m\by_m=0$ as $\mathbf{u}=\mathbf{\tilde C
  v}$.  Eliminating $\mathbf{u}$ between this and eq.~\eqref{eq:vfromu} leads to
\begin{equation}
  \label{eq:detuv}
\hat F(\omega)\equiv  \det(\mathbf{\tilde B\tilde C} - \mathbf{I})=0.
\end{equation}
This is analytic, apart from roundoff error, as we have confirmed numerically via
eq.~\eqref{eq:analyticity}  and by the convergence of the iteration \eqref{eq:secant} to the level
$|\Delta\omega/\omega|\lesssim 10^{-9}$.

\bibliographystyle{mn2e} 
\bibliography{pulse}

\bsp

\label{lastpage}

\end{document}